%% file: eah_fgrp4_paper2_astroph.tex
\documentclass[apj,twocolappendix,numberedappendix]{emulateapj}

\usepackage{lipsum}
\usepackage{afterpage}
\usepackage{graphicx}
\usepackage{apjfonts}
\usepackage{amsmath}
\usepackage{longtable}
\usepackage{multirow}
\usepackage{color}
\usepackage{hyperref}
\usepackage{lineno}
\usepackage{acronym}
\usepackage{pifont}
\DeclareGraphicsExtensions{.pdf}
\shorttitle{The Einstein@Home Gamma-ray Pulsar Survey}
\shortauthors{\sc Wu et al.}

\begin{document}
\title{The Einstein@Home Gamma-Ray Pulsar Survey II. Source Selection, Spectral Analysis and Multi-wavelength Follow-up}
\author{
J.~Wu\altaffilmark{1,2},
C.~J.~Clark\altaffilmark{3,4},
H.~J.~Pletsch\altaffilmark{3,4},
L.~Guillemot\altaffilmark{1,5,6,7},
T.~J.~Johnson\altaffilmark{8}, 
P.~Torne\altaffilmark{1},
D.~J.~Champion\altaffilmark{1}, 
J.~Deneva\altaffilmark{8}, 
P.~S.~Ray\altaffilmark{9}, 
D.~Salvetti\altaffilmark{10},
M.~Kramer\altaffilmark{1,11,12},
C.~Aulbert\altaffilmark{3,4}, 
C.~Beer\altaffilmark{3,4}, 
B.~Bhattacharyya\altaffilmark{13}, 
O.~Bock\altaffilmark{3,4}, 
F.~Camilo\altaffilmark{14}, 
I.~Cognard\altaffilmark{5,6}, 
A.~Cu\'ellar\altaffilmark{3,4}, 
H.~B.~Eggenstein\altaffilmark{3,4}, 
H.~Fehrmann\altaffilmark{3,4}, 
E.~C.~Ferrara\altaffilmark{15}, 
M.~Kerr\altaffilmark{9}, 
B.~Machenschalk\altaffilmark{3,4}, 
S.~M.~Ransom\altaffilmark{16}, S.~Sanpa-Arsa\altaffilmark{17}, 
K.~Wood\altaffilmark{18}
}
\altaffiltext{1}{Max-Planck-Institut f\"ur Radioastronomie, Auf dem H\"ugel 69, D-53121 Bonn, Germany}
\altaffiltext{2}{email: jwu@mpifr-bonn.mpg.de}
\altaffiltext{3}{Albert-Einstein-Institut, Max-Planck-Institut f\"ur Gravitationsphysik, D-30167 Hannover, Germany}
\altaffiltext{4}{Leibniz Universit\"at Hannover, D-30167 Hannover, Germany}
\altaffiltext{5}{Laboratoire de Physique et Chimie de l'Environnement et de l'Espace -- Universit\'e d'Orl\'eans / CNRS, F-45071 Orl\'eans Cedex 02, France}
\altaffiltext{6}{Station de radioastronomie de Nan\c{c}ay, Observatoire de Paris, CNRS/INSU, F-18330 Nan\c{c}ay, France}
\altaffiltext{7}{email: lucas.guillemot@cnrs-orleans.fr}
\altaffiltext{8}{College of Science, George Mason University, Fairfax, VA 22030, resident at Naval Research Laboratory, Washington, DC 20375, USA}
\altaffiltext{9}{Space Science Division, Naval Research Laboratory, Washington, DC 20375-5352, USA}
\altaffiltext{10}{INAF-Istituto di Astrofisica Spaziale e Fisica Cosmica Milano, via E. Bassini 15, I-20133 Milano, Italy}
\altaffiltext{11}{Jodrell Bank Centre for Astrophysics, School of Physics and Astronomy, The University of Manchester, M13 9PL, UK}
\altaffiltext{12}{University of Manchester, Manchester, M13 9PL, UK}
\altaffiltext{13}{National Centre for Radio Astrophysics, Tata Institute of Fundamental Research, Pune 411 007, India}
\altaffiltext{14}{Square Kilometre Array South Africa, Pinelands, 7405, South Africa}
\altaffiltext{15}{NASA Goddard Space Flight Center, Greenbelt, MD 20771, USA}
\altaffiltext{16}{National Radio Astronomy Observatory, Charlottesville, VA 22903, USA}
\altaffiltext{17}{National Astronomical Research Institute of Thailand, 191 Siriphanich Bidg. 2nd Fl. Huay Kaew Rd. Suthep District, Muang, Chiang Mai 50200, Thailand}
\altaffiltext{18}{Praxis Inc., Alexandria, VA 22303, resident at Naval Research Laboratory, Washington, DC 20375, USA}

\begin{abstract} 
\noindent

We report on the analysis of 13 gamma-ray pulsars discovered in the Einstein@Home blind search survey using \textit{Fermi} Large Area Telescope (LAT) Pass 8 data. The 13 new gamma-ray pulsars were discovered by searching 118 unassociated LAT sources from the third LAT source catalog (3FGL), selected using the Gaussian Mixture Model (GMM) machine learning algorithm on the basis of their gamma-ray emission properties being suggestive of pulsar magnetospheric emission. The new gamma-ray pulsars have pulse profiles and spectral properties similar to those of previously-detected young gamma-ray pulsars. Follow-up radio observations have revealed faint radio pulsations from two of the newly-discovered pulsars, and enabled us to derive upper limits on the radio emission from the others, demonstrating that they are likely radio-quiet gamma-ray pulsars. We also present results from modeling the gamma-ray pulse profiles and radio profiles, if available, using different geometric emission models of pulsars. The high discovery rate of this survey, despite the increasing difficulty of blind pulsar searches in gamma rays, suggests that new systematic surveys such as presented in this article should be continued when new LAT source catalogs become available.

\end{abstract} 


\newacro{LAT}[LAT]{\textit{Fermi} Large Area Telescope}
\newacro{Fermi}[\emph{Fermi}]{\emph{Fermi}}
\newacro{VHE}[VHE]{very high energy}
\newacro{PWN}[PWN]{pulsar wind nebula}


\section{Introduction}
\label{s:intro}

Pulsars are rapidly rotating neutron stars with rotational periods ranging from more than 10 seconds to just a few milliseconds. Since their discovery in 1967 \citep{1968Natur.217..709H}, various pulsar surveys have discovered over 2600 pulsars \footnote{ http://www.atnf.csiro.au/research/pulsar/psrcat/}. While the large majority of the known pulsars have been detected in the radio, pulsars are occasionally detected at optical, infrared, UV, X-ray or even gamma-ray frequencies, enabling multi-wavelength studies \citep[see,][for recent examples]{Swiggum2017,Mignani2017}.

During the first eight years of operation, over 200 gamma-ray pulsars have been detected by the \textit{Fermi} Large Area Telescope\footnote{See https://tinyurl.com/fermipulsars for the list of LAT-detected pulsars} \citep[LAT, ][]{Atwood2009ApJ}. The majority of the detected gamma-ray pulsars were first found in radio, either discovered from radio pulsar surveys or targeted radio observations of unassociated LAT sources \citep[i.e. sources with no obvious counterparts at other wavelengths, see, e.g.,][]{Cognard2011ApJ,Keith2011MNRAS,Ransom2011ApJ,Camilo2012ApJ,Guillemot2012MNRAS,Kerr2012ApJ,Barr2013MNRAS,Bhattacharyya2013ApJ,Camilo2015ApJ,Cromartie2016ApJ}. However, a substantial fraction of the gamma-ray pulsars have been discovered by direct, blind searches of the LAT data \citep[e.g.,][]{Abdo2009Sci,SazParkinson2010ApJ,Pletsch+2012-9pulsars,Clark2015ApJJ1906}.

Gamma-ray pulsars found in blind searches are interesting for many reasons. These pulsars are young and energetic with characteristic ages < 3 Myr and spin down power $\dot{E}$ > $10^{33}$ erg s$^{-1}$\citep[see Figure 1 of the second \textit{Fermi} LAT catalog of gamma-ray pulsars, hereafter 2PC, ][]{2PC}. These young energetic pulsars often have timing noise and glitches. This absence of timing coherence makes their pulsations more difficult to find in the low count-rate gamma-ray data acquired over time spans of years. The discovery of PSR J1906+0722 \citep{Clark2015ApJJ1906} demonstrated the ability of the improved semi-coherent blind search technique to detect pulsars even when the data contain timing noise and a substantial glitch. Such blind search methods can reduce the bias against the discovery of young and energetic radio-quiet pulsars in the current pulsar population.


Although the 41 pulsars found in previous blind gamma-ray searches represent a small fraction of the total pulsar population, this increasing population form a very distinct group with extremely faint or undetectable radio emission. Besides the possible detections of J1732$-$3131 \citep{Maan2012_J1732} and Geminga \citep{Maan2015_geminga}, only four gamma-ray discovered pulsars have also been detected in radio, two of them being radio-loud (we follow the convention used in 2PC, i.e., pulsars are considered radio-quiet if their radio flux densities at 1400 MHz, $S_{1400}$, are smaller than $30\ \mu$Jy), J1741$-$2054 and J2032+4127 \citep{Camilo2009ApJ} and the remaining two, J0106+4855 \citep{Pletsch+2012-9pulsars} and J1907+0602 \citep{Abdo2010ApJ}, are considered radio-quiet.

To further increase the number of known gamma-ray pulsars, a new blind search of unidentified LAT sources with gamma-ray emission properties resembling known pulsars was initiated. This search has been conducted on the distributed volunteer computing system Einstein@Home\footnote{https://einsteinathome.org} using the newly improved Pass 8 LAT data. This dataset provides a number of improvements such as better energy reconstruction and better background rejection \citep[see ][]{Atwood2013} therefore increasing its sensitivity. 

Based on their gamma-ray properties, we have selected and searched 118 unassociated LAT sources, resulting in the discovery of 17 pulsars. The results of this search are presented in two papers; Paper~I \citep{PaperI} focused on the search method,  sensitivity and temporal characteristics of the recent pulsar discoveries. In this second paper, we present the source selection scheme, the data preparation process, and detailed gamma-ray analyses and radio follow-up observations of the discoveries. In Section~\ref{s:source_selection}, we describe the method used for selecting gamma-ray sources for the blind search. Section~\ref{s:data_preparation} describes the analysis procedure we followed for preparing the gamma-ray data to be searched for pulsars. Gamma-ray, X-ray and radio follow-up analyses of the newly discovered pulsars are described in Section~\ref{s:follow_up}, and we conclude with a discussion of the properties of the new pulsars. 


\section{Source selection}
\label{s:source_selection}

\subsection{3FGL catalog}
\label{s:3FGL}

The third catalog of LAT sources \citep[hereafter 3FGL,][]{3FGL} lists the properties of 3033 gamma-ray sources detected by the LAT in the first four years of data taking. More than 30\% of the 3FGL sources were unassociated at the time of publication. More than one hundred of these unassociated sources have been demonstrated to be previously unknown pulsars, discovered either in deep targeted radio observations or in blind searches using the LAT data. Due to the observing time and processing resources required for a timing search, identifying which of these sources are most likely to be pulsars has become a task of paramount importance. In contrast to several other classes of gamma-ray sources, pulsars have significant cutoffs in their emission spectra at energies of a few GeV and gamma-ray fluxes that are generally very stable \citep[however see][for a counter-example]{Allafort2013ApJ}; hence the curvature significance\footnote{Significance (in $\sigma$ units) of the fit improvement when assuming a curved spectral type (e.g., PLEC, see Section~\ref{s:pipeline}) instead of a simple power-law for the source of interest. Values greater than 4 indicate significant curvature.} (``Signif\_Curve'', $S_c$) and the variability index\footnote{Index quantifying the variability of a source on a time scale of months. An index larger than 72.44 corresponds to a >99\% confidence probability that the source of interest has a variable flux.} (``Variability\_Index'', VI), which are respectively measures of the curvature of a source's spectrum and of its gamma-ray flux variability, have been successfully applied in previous similar surveys \citep[e.g.,][]{Pletsch+2012-9pulsars}.

We note that only a preliminary version of the 3FGL catalog was available when our survey was initiated. We therefore assessed the pulsar likelihood of the unassociated sources from this preliminary catalog. As a cross-check of our source selection results we have compared the data from the preliminary catalog with those from 3FGL, finding differences in one specific parameter only. These differences are discussed in the next section.

\subsection{Pulsar candidate selection}
\label{s:classification}

Although using $S_{c}$ and VI seems to be enough to identify pulsar candidates, extra care needs to be taken as these two parameters are correlated with the detection significance. A number of groups have developed different schemes for classifying sources, involving machine learning techniques \citep{Lee2012+GMM,Mirabal2012,Saz_Parkinson_2016}. In particular, \citet{Lee2012+GMM} have shown that including the gamma-ray flux as a third dimension in the pulsar classification scheme can directly correct the above-mentioned correlation. Applying the Gaussian Mixture Model (GMM) classification scheme from \citet{Lee2012+GMM}, we used the VI, $S_{c}$ and $F_{1000}$ (gamma-ray flux above 1 GeV) parameters from the catalog to calculate the pulsar likelihood $R_{s}$ for all the sources. A positive log $R_{s}$ indicates that the source is likely to be a pulsar \citep[see][for a detailed discussion]{Lee2012+GMM}. A list of 341 sources with positive logarithmic pulsar likelihood (log $R_{s}$) values and no firm associations with any other astrophysical sources was obtained. 

As mentioned in Section~\ref{s:3FGL}, the list of pulsar candidates was produced by analyzing a preliminary version of the 3FGL catalog. We verified that the characteristics of most of the sources from the preliminary catalog are identical to those from the final catalog. One difference concerns the definition of the spectral curvature Test Statistic (TS), $\textrm{TS}_\textrm{curve}$, listed instead of the curvature significance in the preliminary version of the catalog, $S_{c}= \sqrt{\textrm{TS}_\textrm{curve} \times \textrm{R}_\textrm{syst}}$, where $\textrm{R}_\textrm{syst}$ accounts for systematic uncertainties in the effective area. We verified that using $\textrm{TS}_\textrm{curve}$ instead $S_{c}$ as one of the inputs of the GMM does not affect our classification results. 


\section{Data preparation}
\label{s:data_preparation}

\subsection{The spectral analysis pipeline}
\label{s:pipeline}

One of the main difficulties in blind searches for gamma-ray pulsars is separating background emission from photons originating from the sources of interest. Due to the wide and energy dependent point-spread function of the LAT at low energies\footnote{https://www.slac.stanford.edu/exp/glast/groups/canda/lat\_Performance.htm}, neighboring sources within a few degrees of a given direction can raise the background level in the dataset considered for the search. In the past, blind searches often adopted a so-called ``cookie cutter'' to select photons and increase the signal-to-noise ratio, i.e., they restricted the region of interest (RoI) by selecting events with reconstructed directions found within, say, $\sim 1\arcdeg$ of the considered sky location. Although this technique can efficiently separate source and background photons for some bright pulsars or pulsars in regions of low background contamination, most of the young gamma-ray pulsars are located near the Galactic plane, where the diffuse background emission is strong and where the effectiveness of the cookie cutter selection method decreases. \citet{Kerr2011} mitigated this problem by proposing a photon-weighting technique, which uses information about the spectrum of the targeted source and the instrumental response of the LAT. Probabilities that photons originate from the source can then be calculated, relaxing the need to select narrow sky regions and greatly improving our sensitivity to weak periodic signals. 

Consequently, accurately determining the spectra of the sources we want to search for pulsations is key for calculating photon weights and thereby increasing the signal-to-noise ratio. We assembled a spectral analysis pipeline based on the \texttt{Pointlike} analysis package \citep{Kerr2010}, allowing us to derive the spectral parameters of the search targets, and to assign good photon weights for the selected datasets. We initially considered LAT data recorded between 2008 August 4 and 2014 April 6 for our survey, and included photons recorded until 2015 July 7 after a few tens of sources had been searched (see Sec. \ref{s:reloc}). We used the \textit{Fermi} Science Tools\footnote{http://fermi.gsfc.nasa.gov/ssc/data/analysis/software} to extract Pass 8 Source class events, processed with the \texttt{P8\_SOURCE\_V3} instrument response functions (IRFs). The Science Tools, IRFs and models for the Galactic and extragalactic diffuse gamma-ray emission used here are internal pre-release versions of the Pass 8 data analysis, which were the latest versions available to us when the survey began, The differences in the best-fit parameters are marginal, compared to the analysis with the most recent IRFs. Therefore, the weights as calculated with the old IRFs are also very similar. Specifications of follow-up data analyses are given in Sec. \ref{s:follow_up}. We used \texttt{gtselect} to select photons with reconstructed directions within 8$\arcdeg$ of the 3FGL positions, photon energies $>100$ MeV and zenith angles $< 100\arcdeg$. We only included photons detected when the LAT was operating in normal science mode, and when the rocking angle of the spacecraft was less than $52\arcdeg$. Photons were then binned into 30 logarithmically-spaced energy bins, and with a spatial bin size of 0.1$\arcdeg$.

For each 3FGL target, a spectral model for the sources within the corresponding RoIs was constructed by including all 3FGL sources within 13$\arcdeg$. Spectral parameters of point sources within 5$\arcdeg$ were allowed to vary. A binned maximum likelihood analysis was performed to measure the gamma-ray spectra of the targeted sources, which were modeled with exponentially cut-off power laws (``PLEC'' spectral shapes), of the form:

\begin{equation}
\frac{dN}{dE} = K \left(\frac{E}{\textrm{1\ GeV}}\right)^{-\Gamma} \textrm{exp}\left(-\frac{E}{E_\textrm{cut}}\right),
\label{e:PLEC}
\end{equation}

where $K$ is a normalization factor, $E_\mathrm{cut}$ is the cutoff energy and $\Gamma$ is the photon index. The above expression accurately reproduces the phase-averaged spectral properties of the majority of known gamma-ray pulsars (see, e.g., 2PC). The normalization parameters of the Galactic diffuse emission and the isotropic diffuse background components were left free in the fits. The best-fit source models from the likelihood analysis with \texttt{Pointlike} were used as inputs for \texttt{gtsrcprob}, to determine the probabilities that the selected photons were indeed emitted by our targets. 

In order to verify the goodness of the fits and check for possible issues in the likelihood results, we produced source significance TS maps and plots of the Spectral Energy Distribution (SED) for each analyzed source. For some of the sources, the best-fit cut-off energies were suspiciously high and were in particular much higher than those of known gamma-ray pulsars. These sources have spectra with low curvature, and could potentially be associated with Supernova Remnants (SNRs) or Pulsar Wind Nebulae (PWNe) which are known to have harder spectra than pulsars. For some sources a very high cutoff energy close to the upper bound of 1 TeV used for the fit was found, suggesting low spectral curvature. In some cases the best-fit photon index $\Gamma$ was close to 0. These low photon indices were found for sources with low TS values. We flagged these problematic sources, but included them in the survey despite the abnormal spectral results since we may still be able to detect pulsations from these sources. We note that SEDs for the latter sources were generally consistent with 3FGL results. In addition, for a small number of 3FGL sources our analysis failed to converge, possibly because of complicated sky regions. Those sources were removed from the target list, and will be revisited in the future. As a result, the original target list was trimmed down to 118 sources, which are listed in the table in the Appendix.

We eventually obtained datasets consisting of lists of photon arrival times to be searched for pulsations, photon weights, and spacecraft positions calculated at each photon time, which are necessary to correct the arrival times for Doppler shifts caused by the motion of the telescope with respect to the sources. These datasets were then passed to the blind search algorithm, for searching for new pulsars among our target sources.

\subsection{Relocalization}
\label{s:reloc}

\input{reloc}

Following the first few discoveries (summarized in Section~\ref{s:search_summary}), we noticed that the timing positions of a few pulsars (see Paper~I for the timing positions of the discovered pulsars) were well outside the 95\% confidence regions from the 3FGL catalog. The observed discrepancies could be caused by the fact that 3FGL catalog positions were determined using 4 years of Pass 7 reprocessed data, while we used 5.5 years of Pass 8 data, which have higher angular resolution. To mitigate this discrepancy we relocalized the sources using \texttt{Pointlike}, by varying the sky coordinates of the sources until the maximum likelihood was found. The results of the relocalization analysis for the new pulsars are given in Table~\ref{t:reloc}.

In most cases, the relocalized positions are closer to the pulsar timing positions than the catalog ones. In addition, the 95\% semi-major axes of the relocalized positions are smaller than in 3FGL. Although this implies a smaller number of trials in sky position for the blind search, leading to a greatly reduced overall computational cost, the true pulsar positions may still fall outside of the error ellipses. In some cases, the timing position is found to be out of both the 3FGL error ellipse and the ellipse from our analysis. From the 47$^{th}$ source onwards, we therefore adopted the relocalized positions with three times the 1-$\sigma$ Gaussian uncertainty reported by \texttt{Pointlike} to obtain a more conservative sky coverage, and also extended our dataset by including photons recorded until 2015 July 7 when the relocalization was done. The inaccurate source locations might have resulted from the imperfect Galactic background model.

\subsection{Search summary} 
\label{s:search_summary}

The blind search survey of the sources listed in the appendix, described in detail in Paper I, enabled the discovery of 17 gamma-ray pulsars. \citet{Clark2015ApJJ1906} reported on the discovery of PSR~J1906+0722, an energetic pulsar with a spin frequency of 8.9 Hz which suffered one of the largest glitches ever observed for a gamma-ray pulsar. \citet{ClarkJ1208} later presented the discovery of PSR~J1208$-$6238, a 2.3 Hz pulsar with a very high surface magnetic field and a measurable braking index of about 2.6. Paper I and the present paper report on 13 young, isolated gamma-ray pulsars also found in this survey. The new pulsars have rotational periods ranging from $\sim$79 ms to 620 ms. They are all energetic, with spin-down powers between about $10^{34}$ erg s$^{-1}$ and $4 \times 10^{36}$ erg s$^{-1}$. Among these, PSRs J1057$-$5851 and J1827$-$1446 are the slowest rotators among currently known gamma-ray pulsars. PSR~J1844$-$0346 experienced a very large glitch in mid-2012 (see Paper I for details). In the next sections we describe dedicated follow-up studies of these 13 pulsars. Finally, we note that two more pulsars were found in this survey: PSRs J1035$-$6720 and J1744$-$7619. These two pulsars will be presented in a separate publication (Clark et al., 2017 submitted).


\section{Follow-up Analysis}
\label{s:follow_up}

\subsection{Spectral Analysis}
\label{s:spectral_analysis}

\input{on_pulse}

\input{off_pulse}

\input{pulse_shape}

\begin{figure*}
\epsscale{1.1}
\plotone{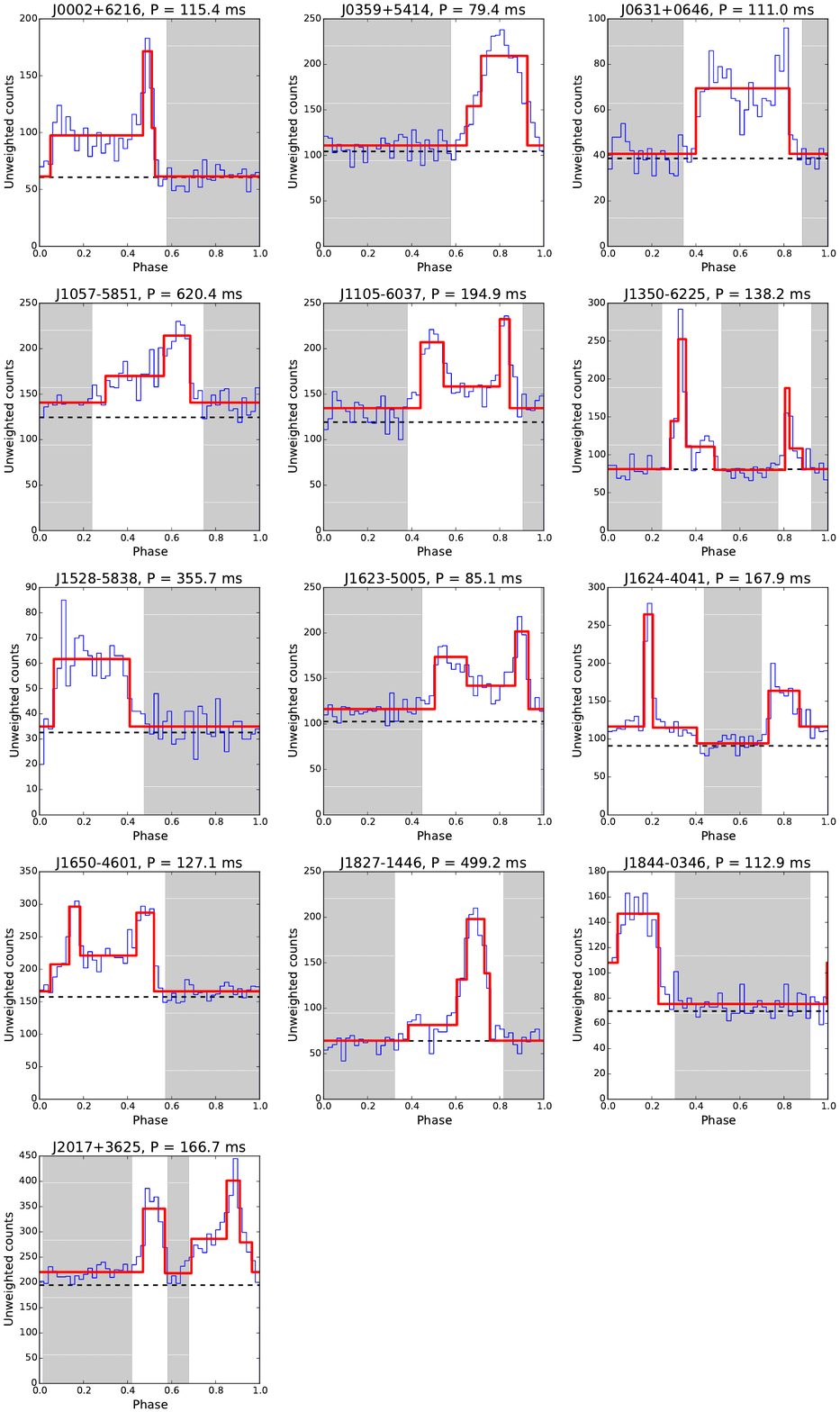}
\caption{Results of the decomposition of gamma-ray pulse profiles into Bayesian blocks, as discussed in Section~\ref{s:spectral_analysis}. Blue histograms represent the pulse profiles, red lines the Bayesian block decompositions, and shaded regions the off-pulse phase intervals determined from this analysis. Dashed black lines represent the estimated background levels, calculated as $B = \sum_i^N \left( 1 - w_i\right)$ where $w_i$ is the weight associated with photon $i$ \citep{Guillemot2012ApJ}. \label{fig:offpulse_def}}
\end{figure*}

\begin{figure*}
\epsscale{1.2}
\plotone{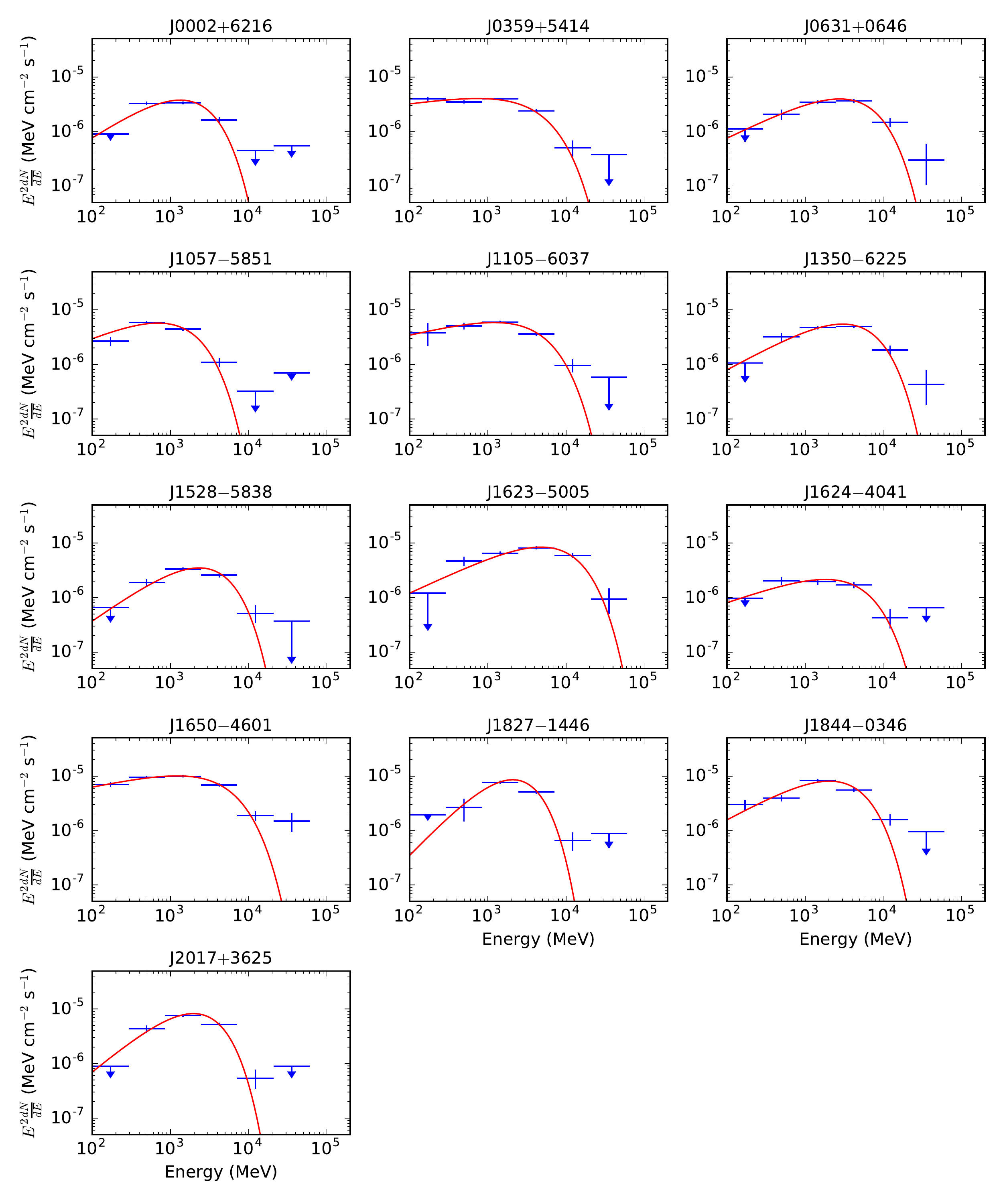}
\caption{Spectral energy distributions for the 13 Einstein@Home pulsars presented in this paper. The best-fit spectral models obtained by analyzing the full energy range are shown as red lines. 95\% confidence upper limits are calculated for energy bins with TS values below 4.\label{fig:SED}}
\end{figure*}

\begin{figure*}
\epsscale{1}
\plotone{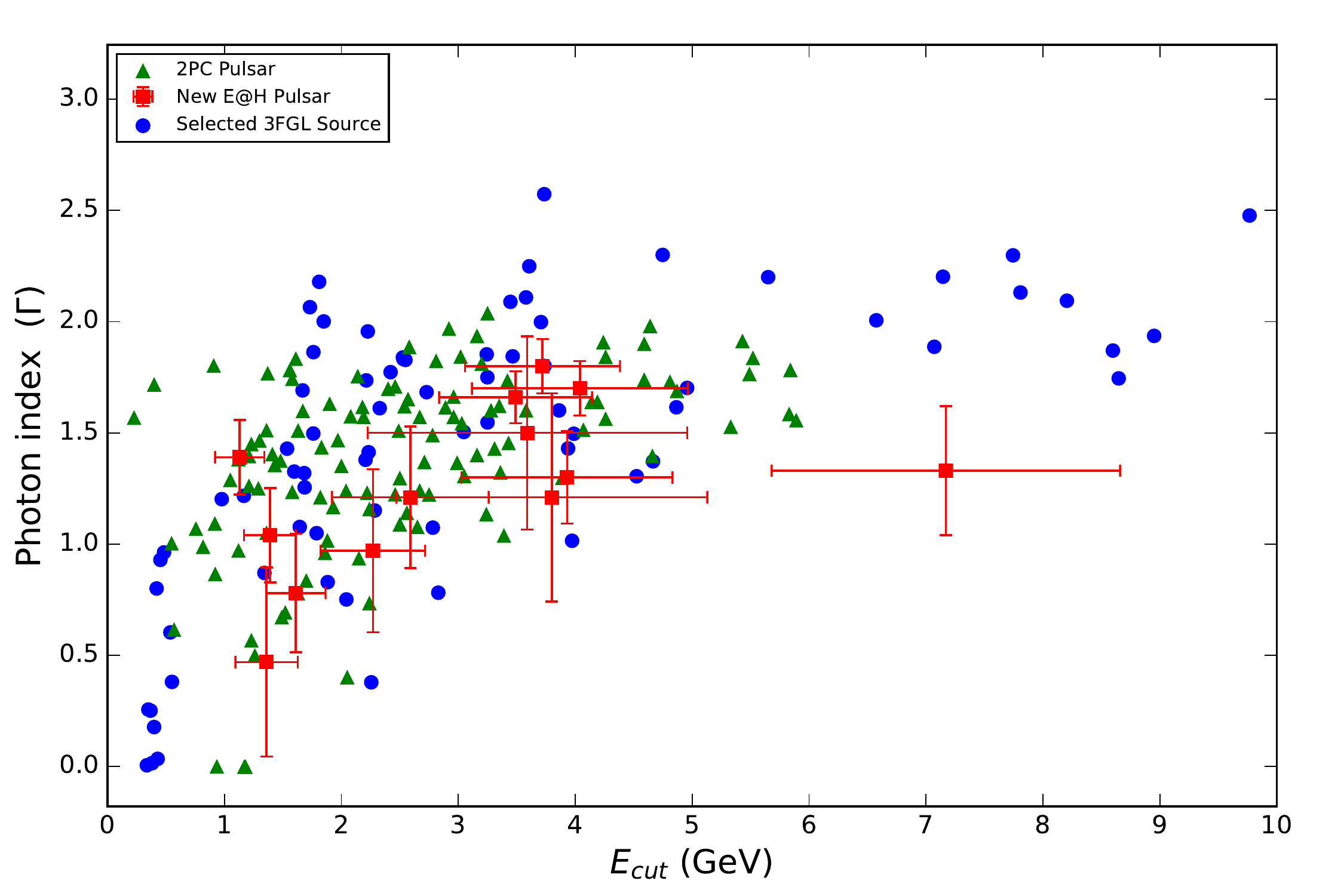}
\caption{Best-fit power-law index $\Gamma$ versus cutoff energy $E_\mathrm{cut}$ for the new pulsars (red squares), other selected 3FGL sources that were searched in our survey (blue circles) and known gamma-ray pulsars from 2PC (green triangles).  3FGL sources with cutoff energies above 10 GeV are not plotted and uncertainties are not displayed, to improve readability.\label{fig:CV}}
\end{figure*}

After selecting the 13 pulsars, we performed dedicated spectral analyses with extended datasets in order to characterize their spectral properties with extra sensitivity. We used updated Pass 8 (P8R2) event selections and updated IRFs for events recorded from 2008 August 4 until 2015 September 9. The sizes of the RoIs around each pulsar were extended to 15$\arcdeg$ to collect more gamma-ray photons for the follow-up analysis, and we selected photon energies $> 100$ MeV and zenith angles $< 90\arcdeg$. The more restrictive zenith angle cut was used to better reject events from the Earth's limb in support of spectral analysis down to 100 MeV. In our dedicated spectral analyses we used the \texttt{gll\_iem\_v06.fits}\footnote{https://fermi.gsfc.nasa.gov/ssc/data/analysis/software/aux/gll\_iem\_v06.fits} map cube and \texttt{iso\_P8R2\_SOURCE\_V6\_v06.txt}\footnote{https://fermi.gsfc.nasa.gov/ssc/data/analysis/software/aux/\\ iso\_P8R2\_SOURCE\_V6\_v06.txt} template for modeling the Galactic diffuse emission and the isotropic diffuse background, to match with the current recommendations \citep{Acero2016ApJS}. The numbers of point sources in the models were increased to include all 3FGL sources within 20$\arcdeg$. The details of the timing analysis using these extended datasets, including the determination of timing and positional parameters, are presented in Paper~I. For the spectral analysis of the 13 pulsars we used the positions obtained from pulsar timing. In order to further minimize contamination from the diffuse background or from neighboring sources, we restricted our datasets to the pulsed part of the pulse profiles. To determine the ``on'' and ``off''-pulse phase regions of the pulse profiles, we selected gamma-ray photons with weights above 0.05 and constructed unweighted pulse profiles, which we then analyzed with the Bayesian Block decomposition method described by \citet{Scargle2013ApJ}. Bayesian Blocks represent a model of time series of events generated by an inhomogeneous Poisson process, involving a sequence of constant flux levels. This method is useful for discriminating random flux changes from real ones, but it is not a physical representation of the pulse profiles. The on- and off-pulse regions are shown in Figure~\ref{fig:offpulse_def}. We selected photons in the on-pulse regions and performed spectral analyses of these restricted datasets. We determined the significance of the spectral cutoff (TS$_\mathrm{cut}$) by comparing the change in log likelihood when using a simple power-law model for the spectra of the pulsars instead of assuming the PLEC model, as follows: TS$_\mathrm{cut} = -2 \log \Delta \mathcal{L}$. The results of the spectral analysis of the on-pulse data are given in Table~\ref{t:params}; the corresponding SEDs are displayed in Figure~\ref{fig:SED}, and the best-fit cutoff energy and power-law index values are shown in Figure~\ref{fig:CV} along with those of 2PC pulsars. 

To search for unpulsed magnetospheric pulsar emission or emission from a putative PWN associated with the pulsar we conducted analyses of the off-pulse phases of the datasets. Point-like test sources were added to the spectral models at the locations of the pulsars, and the spectral properties of these sources were determined by running new likelihood analyses. We alternatively assumed a simple power-law model and a PLEC model for the test sources, in order to test for spectral curvature. Significant off-pulse emission was detected for PSRs~J1623$-$5005, J1624$-$4041 and J2017+3625, with evidence for spectral curvature suggestive of magnetospheric emission from the pulsars, as can be seen from Table~\ref{t:offpulsefit}. Such off-pulse pulsar emission is not atypical for known gamma-ray pulsars (see, e.g., 2PC); nevertheless, small, un-modeled spatial fluctuations in the bright diffuse background emission could also account for this emission. Detailed analyses with extended datasets and comparisons of the best-fit spectral parameters with those of other known gamma-ray pulsars with off-pulse emission are necessary to firmly establish PSRs~J1623$-$5005, J1624$-$4041 and J2017+3625 as pulsars exhibiting gamma-ray emission at all phases. The on-pulse emission was then re-fitted with the addition of sources detected in the off-pulse region scaled to the on-pulse interval with the normalization and spectral parameters fixed.

We characterized the pulse profiles displayed in Figure~5 of Paper I by fitting the weighted profiles to Gaussian or Lorentzian profiles, depending which gave a higher log likelihood. The derived peak multiplicities and gamma-ray peak separations are reported in Table~\ref{t:pulse_shape_parameters}. Most of the new pulsars show double-peaked profiles, with well-separated components that are typical of young gamma-ray pulsar light curves (see 2PC). Two of the 13 newly-discovered pulsars, PSRs~J0002+6216 and J0631+1036, are detected in the radio band (see Section~\ref{s:radio}). For these pulsars we measured the phase offset between the radio peak and the first gamma-ray peak.

\subsection{Radio counterpart searches}
\label{s:radio}

\input{radio_followup_def}

\input{radio_followup_v3}

\begin{figure*}
\epsscale{1}
\plottwo{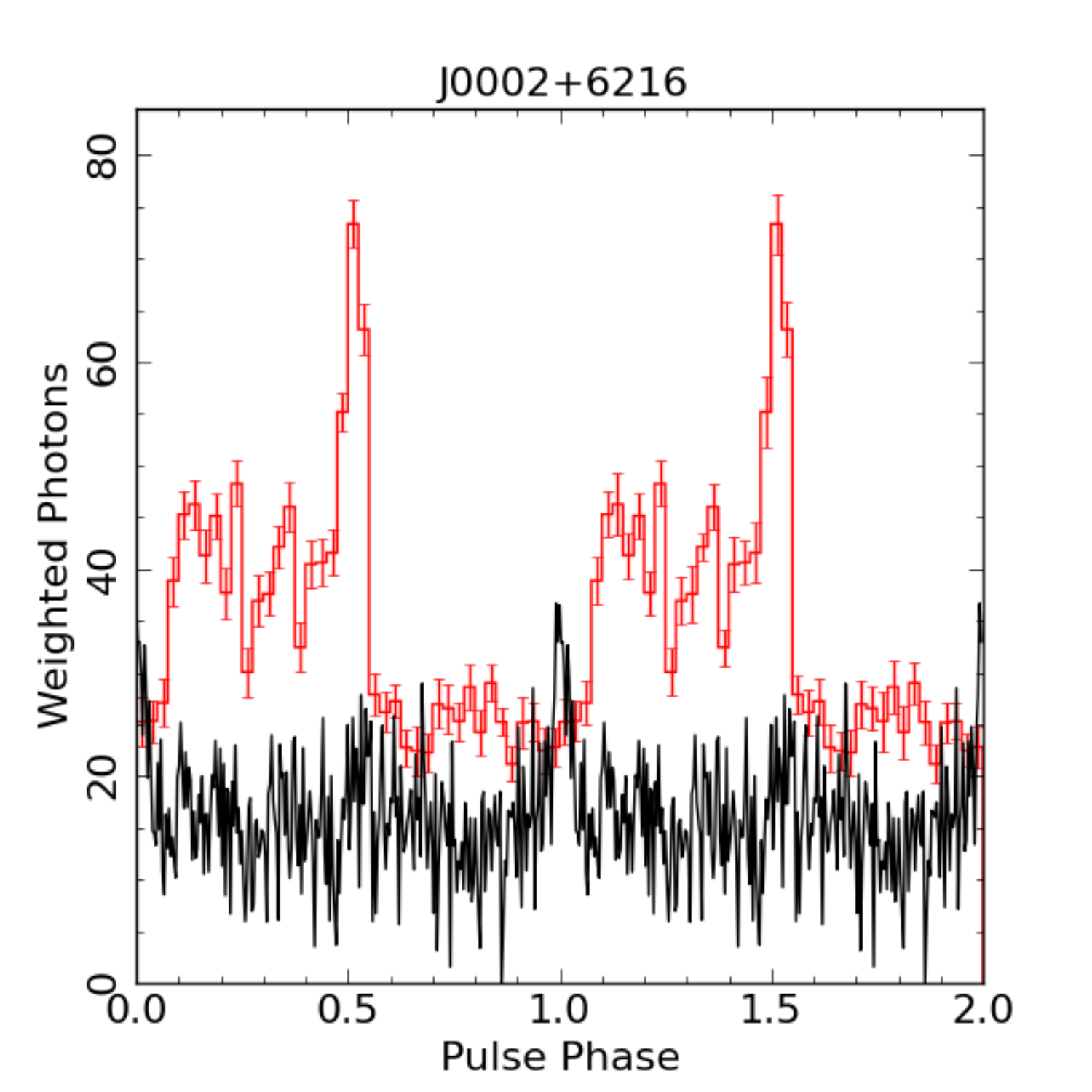}{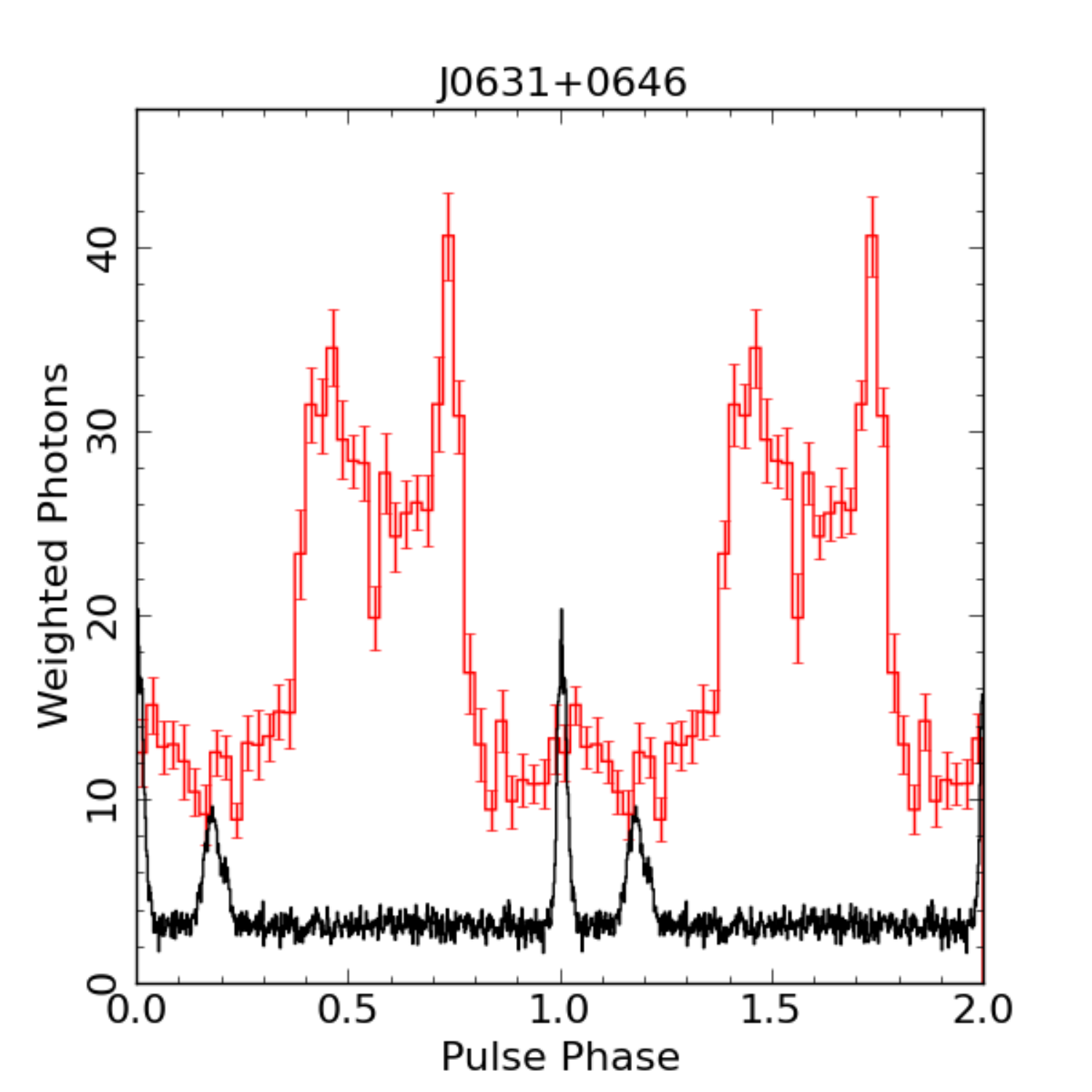}
\caption{Radio and gamma-ray pulse profiles for PSRs~J0002+6216 (left) and J0631+0646 (right). Two complete cycles are shown for clarity. Weighted LAT gamma-ray pulse profiles (in red) were produced by selecting photons with weights greater than 0.05. Radio profiles (in black) correspond to 1.4 GHz observations made with the Effelsberg telescope for J0002+6216 and the Arecibo telescope for J0631+0646.\label{fig:radio_profile} Uncertainties in DM converted to uncertainties in the phase offset between the radio and gamma-ray peaks correspond to $\sim$1\% of the rotational periods.}
\end{figure*}

The new pulsars were searched for radio pulsations by reanalyzing archival observations from previous targeted radio surveys of \textit{Fermi} LAT unassociated sources, or by conducting new dedicated observations. Because we have timing parameters for the new pulsars, the only parameter to search for when analyzing the radio observations is the Dispersion Measure (DM), a quantity representing the integrated column density of free electrons along the line of sight to the pulsars, causing radio waves to arrive at different times depending on the frequency. Radio observations were therefore folded at the periods determined from the gamma-ray timing (see Paper~I) and searched in DM values only, resulting in a reduced number of trials compared to a typical radio pulsar search. 

The list of telescopes and backends used is given in Table~\ref{t:radio_follow_def}. For each observing configuration we give the gain $G$, the central frequency, the frequency bandwidth $\Delta F$, the sensitivity degradation factor $\beta$, the number of polarizations $n_p$, the half width at half maximum of the radio beam (HWHM) and the receiver temperature $T_\mathrm{rec}$. Table~\ref{t:radio_follow_up} lists the radio observations processed in our follow-up study. Sensitivities were calculated using the modified radiometer equation given in \citet{lorimer2005handbook}:

\begin{equation}
S_\mathrm{min} = \beta \dfrac{5 T_\mathrm{sys}}{G \sqrt{n_p t_\mathrm{int} \Delta F}} \sqrt{\dfrac{W}{P - W}}
\end{equation}

\noindent where a value of 5 is assumed for the threshold signal-to-noise ratio for a detection, $T_\mathrm{sys} = T_\mathrm{rec} + T_\mathrm{sky}$, $t_\mathrm{int}$ is the integration time, $P$ is the rotational period and $W$ is the pulse width, assumed to be $0.1 \times P$. The quantity $T_\mathrm{sky}$ is the temperature from the Galactic synchrotron component, estimated by scaling the 408 MHz map of \citet{haslam82} to the observing frequency, assuming a spectral index of $-2.6$. For some observations the pointing direction was offset from the actual sky location of the pulsar. In those cases the flux density limit $S_\mathrm{min}$ as calculated using Equation (2) was divided by $e^{-\left(\theta / \mathrm{HWHM}\right)^2 / 1.5}$, where $\theta$ is the offset. For the majority of pulsars we failed to detect pulsations in the radio data and placed limits on their radio flux densities.  

For two pulsars, PSRs~J0002+6216 and J0631+0646, the analysis resulted in the detection of significant radio pulsations. PSR~J0002+6216 was detected in a 2-hr observation conducted at 1.4 GHz with the Effelsberg radio telescope, with a DM of 218.6(6) pc cm$^{-3}$. PSR~J0631+0636 was detected with Arecibo at 327 MHz and at 1.4 GHz in $\sim$70~min observations, and was also seen with Effelsberg at 1.4 GHz during a 2-hr follow-up observation. The best determined DM value from the Arecibo 327 MHz observation was 195.2(2) pc cm$^{-3}$. Phase-aligned radio and gamma-ray pulse profiles for PSRs~J0002+6216 and J0631+0646 are displayed in Figure~\ref{fig:radio_profile}. In both cases the gamma-ray emission is seen to lag the weak radio emission, as commonly observed in other radio and gamma-ray pulsars and suggesting radio and gamma-ray emissions having different magnetospheric origins (see, e.g., 2PC). 

\subsection{Pulse profile modeling}
\label{s:modeling}

\input{profile_fitting}

Using photons selected within 5$\arcdeg$ radius around the pulsars, we constructed weighted counts pulse profiles with 90, 60, or 30 bins if the weighted H-test TS, the statistical test for pulsation significance \citep{deJager+1989}, for a given pulsar was $\geq$1000, between 100 and 1000, or $<$100, respectively. For PSRs~J0002+6216 and J0631+0646 with radio detections, we re-binned the radio pulse profiles to have the same number of bins as the corresponding gamma-ray profile. We performed likelihood fits, minimizing $-\ln\mathcal{L}$ where $\mathcal{L}$ is the likelihood value, of the gamma-ray pulse profile or the combination of the radio and gamma-ray pulse profiles,  of all 13 pulsars using the geometric simulations and fitting technique of \citet{Johnson2014}.

Following \citet{ClarkJ1208}, we used simulations with $P$ = 100 ms and $\dot{P}$ = 10$^{-15}$ s s$^{-1}$ and constructed likelihood values using a $\chi^{2}$ statistic.  Each pulsar was fit using the outer gap \citep[OG, e.g.,][]{Cheng86} and the slot gap \citep[e.g.,][]{MH03,MH04} models, where we used the two-pole caustic model \citep[TPC,][]{DR03} as a geometric representation of the slot gap.  For both models we use the vacuum retarded dipole solution for the magnetic field geometry \citep{Deutsch55}.  The simulations were produced with 1$\arcdeg$ resolution in both the magnetic inclination angle ($\alpha$) and observer angle ($\zeta$) and a resolution of 1\% of the polar cap opening angle in emitting and accelerating gap widths. For radio simulations, we assumed a frequency of 1400 MHz with the conal geometry and emission altitude of \citet{Story07}.

The best-fit results for all but PSR J0631+0646 are given in Table~\ref{tab:models}; estimated uncertainties are quoted at the 95\% confidence level but note that systematic error estimates from the fitting method \citep[see][]{Johnson2014} and/or from fitting only the gamma-ray profiles \citep{Pierbattista2015} could be as large as 10$\arcdeg$.  \citet{Johnson2014} noted that it was necessary to renormalize the $\Delta\ln\mathcal{L}$ surface, making the best fit approximately correspond to a reduced $\chi^{2}$ value of 1, in order to have more realistic confidence contours.  In some cases, however, we found that this renormalization was unnecessary, either having no effect on the estimated uncertainties or shrinking them.  We denote the pulsars for which we did not renormalize the likelihood surface with a $\dag$ in column 1 of Table~\ref{tab:models}.  For each model, we also estimated the beaming fraction $f_{\Omega}$ \citep[as defined, e.g., in][]{Watters09,VHG09} for the best-fit geometry, used when calculating the gamma-ray luminosity.

For each pulsar with no radio detection, we examined the simulated radio sky map at 1400 MHz, and evaluated the model predictions for the best-fit geometry, in regard to expected radio loudness; the predictions are indicated in columns 6 and 11 of Table~\ref{tab:models}.  The model predictions are: `L' for radio-loud, meaning the predicted geometry has the radio cone clearly and strongly intersecting our line of sight; `F' for radio-faint, meaning the predicted geometry has our line of sight either narrowly missing the cone or clipping the very edge suggesting only weak emission would be detected; and `Q' for radio-quiet, meaning the predicted geometry has our line of sight clearly missing the radio cone.  The radio-faint sources are of particular interest as searches at frequencies lower than 1400 MHz, where the cone is predicted to be larger \citep[e.g.,][]{Story07}, may yield detections. Following \citet{Johnson2014}, we conservatively consider one model to be significantly favored over another, for a given pulsar, if the $\ln\mathcal{L}$ value is greater by at least 15; however, in some cases the best-fit geometry for the TPC model clearly predicts a radio-loud pulsar where none has been detected and we therefore claim the OG model is favored, regardless of the $\Delta\ln\mathcal{L}$ value.  In particular, this is the case for PSRs~J0359+5414, J1528$-$5838, and J1827$-$1446.  For J1350$-$6225, both the TPC and OG model predict a radio-loud pulsar, with a near-orthogonal rotator viewed near the spin equator, either casting doubts on the models or raising questions concerning the non-detection. In modeling the ``radio-quiet'' pulsars in 2PC, \citet{Pierbattista2015} similarly found some solutions where the line of sight was near enough to the magnetic axis that we might expect to intersect the radio emission cone.  These authors used a different fitting technique but similar simulations.  This may further suggest that our results regarding the aforementioned pulsars point to issues with the models and not with the non-dections in radio. 

Our joint gamma-ray and radio fits of PSR~J0631+0646 did not produce acceptable results: the standard approach tended to ignore the radio data.  Following \citet{Johnson2014}, we decreased the radio uncertainty value in order to increase its importance in the likelihood but this proved ineffective, leading to fits that ignored the gamma-ray data.  Under the assumption that the difficulty was in matching the observed phase lag between the radio peak and the gamma-ray peaks, we followed \citet{GuillemotJ0737} in allowing the phase of the magnetic pole in the radio and gamma-ray simulations to be different by as much as 0.1  \citep[following realistic simulations of the pulsar magnetosphere by][which suggested an offset of the low-altitude polar gap from the outer magnetosphere by up to this amount]{Kala2012}.  These new fits were, similarly, unsatisfactory.  We investigated relaxing the maximum phase offset condition and found more acceptable fits with offsets of $\sim$0.3 in phase for both the TPC and OG models. The maximum phase offset of 0.1 is inferred from \citet{Kala2012} by comparing predicted light curves from the vacuum retarded dipole geometry to models with increasing conductivity and finally full force-free models. It seems implausible that this offset could be a factor of 3 larger than predicted in the force-free simulations. Our different attempts to model the radio and gamma-ray profiles jointly being unsuccessful we do not report modeling results for PSR~J0631+0646 in Table~\ref{tab:models}. New approaches for modeling this pulsar's emission geometry are needed. For instance, based on the work of \cite{Kalapotharakos2014ApJ79397K}, it is possible that gamma-ray emission from the current sheet outside the light cylinder could explain the extra phase lag for PSR~J0631+0646 as their simulations did tend to show larger radio to gamma-ray phase lags.

\subsection{Luminosity, distance and gamma-ray efficiency}
\label{s:luminosity}

The fraction of their energy budgets that pulsars convert into gamma-ray radiation is a key question for understanding pulsar emission mechanisms. This requires converting the measured energy flux in gamma rays $G_{100}$ (see Table~\ref{t:params} for the values) into the gamma-ray luminosity, with the relation $L_\gamma = 4 \pi f_\Omega G_{100} d^2$ where $d$ is the distance. As discussed in Section~\ref{s:radio}, most of the 13 Einstein@Home pulsars considered in this study are undetected in radio. For these pulsars we therefore cannot use the DM to infer distances, e.g., using the NE2001 model of free electrons in the Galaxy \citep{Cordes2002}. We can however calculate ``heuristic'' distances, $d_h$, and luminosities, $L^h_\gamma$, as follows:

\begin{equation}
d_{h} = \sqrt{L^h_\gamma/4\pi G_{100}},
\end{equation}

\noindent where

\begin{equation}
L^h_\gamma= \sqrt{\dot{E}/10^{33} \textrm{erg}~\textrm{s}^{-1}} \times 10^{33}~\textrm{erg}~\textrm {s}^{-1},
\end{equation}

\noindent i.e., assuming that the gamma-ray luminosity scales as $\sqrt{\dot E}$ for these young pulsars (see 2PC) and assuming a typical geometrical factor $f_\Omega$ of 1. Heuristic distances for the 13 pulsars are given in Table~\ref{t:pulse_shape_parameters}. In most cases the values suggest that the pulsars lie at small or intermediate distances, as is also the case for the majority of known gamma-ray pulsars. 

From the radio detections of PSRs~J0002+6216 and J0631+0646 we could determine DM values and use the NE2001 model to extract the DM distances given in Table~\ref{t:pulse_shape_parameters}. For both pulsars the NE2001 distance is very large. The distance for PSR~J0002+6216 of 7.7 kpc leads to a gamma-ray efficiency $\eta = L_\gamma / \dot E$ of about 120\%. For PSR~J0631+0646 a conversion efficiency of 100\% is found for a distance of about 6.7 kpc. The NE2001 model therefore probably underestimates the density of free electrons along the lines of sight to these pulsars. Interestingly, the recently-published model for the distribution of free electrons in the Galaxy of \citet{ymw17} finds DM distances of 6.3 kpc and 4.6 kpc for PSR~J0002+6216 and J0631+0646, respectively. The latter distance values lead to realistic efficiency estimates below 100\% (81\% and 90\% respectively).


\subsection{X-ray counterpart searches}
\label{s:x-ray}

We re-analyzed archival X-ray observations to search for counterparts to the new gamma-ray pulsars and to characterize their X-ray spectra. All our targets except PSR~J1827$-$1446 have adequate coverage by at least one of the major contemporary observatories operating in the soft X-ray band: {\it Swift} \citep{Burrows2005SSRv}, {\it XMM-Newton} \citep[]{Strder2001A&A,Turner2001AA} and {\it Chandra} \citep{Garmire2003SPIE}. The X-ray coverage ranges from few-ks shallow snapshots with {\it Swift} to orbit-long, deep observations by {\it Chandra} and {\it XMM-Newton}. Almost all the detected pulsars have been observed by {\it Swift} as part of a systematic survey of the gamma-ray error boxes of the unidentified {\it Fermi} LAT sources \citep{Stroh2013ApJS}. 

We reduced and analyzed the {\it XMM-Newton} data through the most recent release of the {\it XMM-Newton} Science Analysis Software (SAS) v15.0. We performed a standard data processing, using the {\tt epproc} and {\tt emproc} tools, and screening for high particle background time intervals following \citet{Salvetti2015ApJJ2039}. For the {\it Chandra} data analysis we used the Chandra Interactive Analysis of Observations (CIAO) software version 4.8. We re-calibrated event data by using the {\tt chandra\_repro} tool. {\it Swift} data were processed and filtered with standard procedures and quality cuts\footnote{More detail in: http://swift.gsfc.nasa.gov/docs/swift/analysis/} using FTOOLS tasks in the HEASOFT software package v6.19 and the calibration files in the latest Calibration Database release.

We performed a standard data analysis and source detection in the 0.3$-$10 keV energy range of the {\it XMM-Newton}-EPIC, {\it Chandra}-ACIS and {\it Swift}-XRT observations (e.g., \citealt{Salvetti2015ApJJ2039} and \citealt{Marelli2015ApJ}). We preferred the XMM and Chandra observatories if the same field has been observed because of the better performance in terms of effective area and spatial resolution. For each of the X-ray counterparts we performed a spectral analysis using XSPEC v12.9. After extracting response matrices and effective area files, we extracted X-ray fluxes by fitting the spectra with a power-law (PL) model using either a $\chi^2$ or the C-statistic \citep{Cash1979ApJ} in the case of low counts ($<$ 100 photons) and negligible background. For sources characterized by low statistics (typically $\leq$ 30 photons), we fixed the column density to the value of the Galactic N$_{\textrm{H}}$ integrated along the line of sight (Dickley \& Lockman 1990) and scaled for the heuristic distance and, if necessary, set the X-ray PL photon index ($\Gamma_{\textrm{X}}$) to 2. All quoted uncertainties on the spectral parameters are reported at the 1$\sigma$ confidence level. For each pulsar we computed the corresponding gamma-ray-to-X-ray flux ratio. As reported in \cite{Marelli2015ApJ}, this could give important information on the nature of the detected pulsar. Finally, for all undetected ones, we computed the 3$\sigma$ X-ray detection limit based on the measured signal-to-noise ratio, assuming a PL spectrum with $\Gamma_{\textrm{X}}=2$ and the integrated Galactic N$_{\textrm{H}}$, scaled for the heuristic distance. The detailed results of these analyses are reported in Table~\ref{tab:1}.

Out of the 13 gamma-ray pulsars, we detected a significant X-ray counterpart for six. PSR J0002+6216, PSR J1105$-$6037 and PSR J1844$-$0344 were detected with {\it Swift}-XRT. These sources are listed in the First Swift XRT Point Source (1SXPS) Catalogue \citep{Evans2014ApJS} as 1SXPS J000257.6+621609, 1SXPS J110500.3$-$603713 and 1SXPS J184432.9$-$034626, respectively. These sources are located at ($\alpha$, $\delta$) (J2000) = (0.7404$\arcdeg$, +62.2692$\arcdeg$), (166.2515$\arcdeg$, $-$60.6203$\arcdeg$) and (281.1371$\arcdeg$, $-$3.7740$\arcdeg$) with 90\% confidence error circles of 4.9$''$, 6.4$''$ and 2.7$''$. Owing to the long {\it Chandra} exposure time, we clearly detected both the pulsar and the associated nebula of PSR J0359+5414. The pulsar is located at ($\alpha$, $\delta$) = (59.8586$\arcdeg$, +54.2486$\arcdeg$) with a 90\% confidence error circle of 1$''$. The nebula is approximately elliptical, with semi-major and semi-minor axes of $\sim15''$ and $\sim7''$, respectively, roughly centered on the pulsar position. The nebula is well fitted by an absorbed PL model with photon index equal to 1.4$\pm$0.2 and unabsorbed flux in the 0.3$-$10 keV energy band of (1.3$\pm$0.3)$\times10^{-14}$ erg cm$^{-2}$ s$^{-1}$. We also detected the counterpart for PSR J2017+3625 at  ($\alpha$, $\delta$) = (304.4827$\arcdeg$, 36.4189$\arcdeg$), with a 2$''$ error, from analysis of a {\it Chandra} observation. From {\it XMM-Newton} data we detected two possible counterparts for PSR J1624$-$4041 at $\sim13''$ from the gamma-ray pulsar position. The two plausible X-ray counterparts are located at ($\alpha$, $\delta$) = (246.0372$\arcdeg$, $-$40.6931$\arcdeg$) and (246.0459$\arcdeg$, $-$40.6899$\arcdeg$) with both a 0.8$''$ statistical plus 1.5$''$ systematic error. We report both counterparts in Table~\ref{tab:1}, as {\it src1} and {\it src2}, respectively.

\input{x-ray}

\section{Discussion}
\label{s:discussion}

A total of 17 gamma-ray pulsars have been discovered among the 118  3FGL sources we have selected for the search, based on their gamma-ray emission properties being suggestive of pulsar emission. The high discovery rate of about 15\% is comparable to that of previous similar surveys, of $\sim$ 8--12\% \citep{Abdo2009Sci,Pletsch+2012-9pulsars,Pletsch2013eh} even though we are searching fainter and fainter LAT sources. The improved semi-coherent blind search technique, the new Pass 8 LAT data, and the improved source selection and localization likely played an important role in the success of the survey. It is interesting to note that a number of sources in our list had already been searched for pulsations in the past. For example, comparing the sky locations searched in our survey with those analyzed in previous Einstein@Home or \textit{Atlas} surveys \citep{Pletsch+2012-9pulsars,Pletsch2013eh}, we find that about 27\% (32 / 118) of our sources had already been searched, and 11 of these have now been found to be gamma-ray pulsars. The multiple improvements in our new gamma-ray blind survey enumerated above likely explain the detections of these pulsars. Similarly, seven of the new discovered pulsars (PSRs J0002+6216, J0631+0646, J1035$-$6720, J1057$-$5851, J1105$-$6037, J1623$-$5005 and J1624$-$4041) fall below the sensitivity limit of the previously used search algorithm (see Section 5.1 of Paper I for more details).

Comparing our target list with the best pulsar candidates from \citet{Saz_Parkinson_2016} who also used machine learning techniques for classifying 3FGL unassociated sources, we find a relatively high overlap of about 60\%. Interestingly, PSRs J0631+0646 and J1827$-$1446, discovered in our survey, do not appear in their list. In the case of PSR~J0631+0646 this could be caused by the possible association with a nearby supernova remnant, while for PSR~J1827$-$1446 the source detection significance of $\sim 9.1 \sigma$ is simply under the 10$\sigma$ threshold set by \citet{Saz_Parkinson_2016} for constructing their list. The good overlap between the two target lists makes us confident that we have selected and searched 3FGL sources likely powered by unknown pulsars.

The spectral properties (photon indices $\Gamma$ and cutoff energies $E_\mathrm{cut}$) for the surveyed sources, for the 13 new gamma-ray pulsars and for pulsars from the 2PC catalog are displayed in Figure~\ref{fig:CV}. The photon indices and cutoff energies of the new Einstein@Home pulsars are very similar to those of 2PC pulsars, a natural consequence of the source selection procedure described in Section~\ref{s:classification}. This is confirmed by a Kolmogorov-Smirnov test, which finds a $\sim$90\% probability that the two samples are drawn from the same parent distribution. The GMM algorithm used for classifying 3FGL sources therefore seems to have efficiently selected pulsar candidates among unassociated sources, which is further supported by the fact that $\sim$ 80\% of the discovered pulsars were found in the top half of Table~\ref{t:source_list}. The gamma-ray fluxes of the new pulsars are generally lower than those of 2PC pulsars found in blind searches, also unsurprisingly. 

Possible reasons for the non-detections of pulsars in other 3FGL unassociated sources listed in Table~\ref{t:source_list} are that these sources could be pulsars with low pulse fractions or broad gamma-ray pulse profiles, for which the sensitivity of the search algorithm is lower (see Section 3 of Paper I for a detailed discussion of the search sensitivity). They could also be pulsars with high timing noise, or they could be millisecond pulsars in binary systems. A number of sources in our list were indeed recently identified as candidate binary MSPs after we started our search: for instance, 3FGL J0212.1+5320 \citep{Li2016ApJJ0212}, 3FGL J0744.1$-$2523 \citep{Salvetti2017J0744} and 3FGL J2039.6$-$5618 \citep{Salvetti2015ApJJ2039}. The discovery of pulsars in binary systems in gamma rays requires initial guesses of the orbital parameters, from, e.g., optical or X-ray observations \citep[see, e.g.,][]{pletsch2012Sci}. If all searched sources are indeed gamma-ray pulsars, then we would expect a good number of them to be in binaries, based on the 2PC pulsar population. 

As can be seen from Figure~\ref{fig:CV} and Table~\ref{t:source_list}, a number of sources included in our survey had $\Gamma$ and/or $E_\mathrm{cut}$ parameters higher than those of 2PC pulsars. These sources were selected by the GMM based on their low variability and moderate curvature indices. Although the gamma-ray emission properties of these sources seem different from those of 2PC pulsars \textit{a priori}, we included them in the survey for completeness but failed to find new pulsars in any of them. One possibility for the future would be to train the GMM not to select these peculiar sources, or to continue searching in order not to miss pulsars with large spectral index and/or cutoff values. In any case, half of the searched sources from Table~\ref{t:source_list} have $\Gamma$ and $E_\mathrm{cut}$ parameters resembling those of 2PC pulsars, and are thus still prime targets for pulsation searches. 

Of the 13 new pulsars reported in this article, only two have been detected in radio. The deep radio follow-up observations conducted as part of this project placed tight constraints on the flux densities of the undetected pulsars. Only six young pulsars among the 54 discovered in blind searches of the LAT data have so far been detected as radio pulsars. The many non-detections in radio are not surprising, given that past radio pulsar surveys have covered the entire sky with moderate sensitivity \citep[see, e.g.,][]{Cordes2006ApJ,Keith2010MNRAS,Barr2013MNRAS,Boyles2013ApJ,Deneva2013ApJ}. PSRs J0002+6216 and J0631+0646, both detected in radio, are however perfect examples of pulsars with low radio flux densities that would be missed in the short integration times of traditional radio pulsar surveys. The LAT was therefore crucial for the discovery of all these young pulsars, and blind search surveys are clearly key for completing the population of young and energetic gamma-ray pulsars. The discovery of these 13 pulsars with Einstein@Home brings the total number of non-recycled gamma-ray pulsars to 112, of which $\sim$ 54\% are radio-loud. The fraction thus remains similar to that reported in 2PC. As the \textit{Fermi} mission continues it will be interesting to see how this fraction evolves, as it is a powerful discriminant of pulsar emission models. 

Recently, \citet{Ajello2017} released a catalog of resolved point sources in a $\textrm{40}\arcdeg \times \textrm{40}\arcdeg$ region around the Galactic center direction. By selecting spectrally curved sources and comparing the spectral energy distributions of these point sources with those of a large sample of 3FGL sources, they could identify pulsar-like candidates from these new Galactic bulge sources. These sources are also prime targets for future blind pulsation searches.

\section{Conclusions}

Using information from a preliminary version of the 3FGL catalog of \textit{Fermi} LAT sources, we have selected 118 targets with pulsar-like emission properties. We produced Pass 8 LAT datasets for each of the sources, and these datasets were then searched for pulsations with a multi-stage blind search algorithm, utilizing the volunteer computing system Einstein@Home. This survey led to the discovery of 17 pulsars, of which 13 are presented in this article, and the other pulsars have been or will be published elsewhere. 

On-pulse and off-pulse gamma-ray spectral analyses were conducted for each of the new pulsars.The gamma-ray emission properties of the 13 newly discovered pulsars reported in this paper are similar to those of other young gamma-ray pulsars, such as those from the 2PC catalog.  Radio follow-up observations were carried out, resulting in the detections of two of them with low radio flux densities. The pulse profiles of the 13 new pulsars were fit using the TPC and OG models. For some of the pulsars, radio emission is predicted by the models but is still undetected in follow-up or archival observations.

The increased sensitivity of the blind search algorithm, the improved Pass 8 LAT data, and improved source selection and relocalization pipeline enabled us to maintain a relatively high detection rate, compared to previous similar surveys. Nevertheless, for a number of the 3FGL sources with clear pulsar-like properties selected for the search, we were unable to find a pulsar. These sources remain excellent targets for future searches. New systematic surveys such as the one presented in this paper and in Paper I are warranted, and so are blind searches for millisecond pulsars in binary systems, which at the moment can only be searched using external constraints on the orbital parameters from observations at other wavelengths.

\acknowledgements

This work was supported by the Max-Planck-Gesellschaft (MPG), by the Deutsche Forschungsgemeinschaft(DFG) through an Emmy Noether research grant PL 710/1-1 (PI: Holger J. Pletsch), and by NSF award 1104902.

The authors also gratefully acknowledge the support of the many
thousands of Einstein@Home volunteers, without whom this search would
not have been possible.

The \textit{Fermi} LAT Collaboration acknowledges generous ongoing support from a number of agencies and institutes that have supported both the development and the operation of the LAT as well as scientific data analysis. These include the National Aeronautics and Space Administration and the Department of Energy in the United States, the Commissariat \`a l'Energie Atomique and the Centre National de la Recherche Scientifique / Institut National de Physique Nucl\'eaire et de Physique des Particules in France, the Agenzia Spaziale Italiana and the Istituto Nazionale di Fisica Nucleare in Italy, the Ministry of Education, Culture, Sports, Science and Technology (MEXT), High Energy Accelerator Research Organization (KEK) and Japan Aerospace Exploration Agency (JAXA) in Japan, and the K.~A.~Wallenberg Foundation, the Swedish Research Council and the Swedish National Space Board in Sweden. 

Additional support for science analysis during the operations phase is gratefully acknowledged from the Istituto Nazionale di Astrofisica in Italy and the Centre National d'\'Etudes Spatiales in France. This work performed in part under DOE Contract DE-AC02-76SF00515.

Fermi research at NRL is supported by NASA. The Parkes radio telescope is part of the Australia Telescope which is funded by the Commonwealth Government for operation as a National Facility managed by CSIRO. We thank our colleagues for their assistance with the radio timing observations. The Robert C. Byrd Green Bank Telescope (GBT) is operated by the National Radio Astronomy Observatory, a facility of the National Science Foundation operated under cooperative agreement by Associated Universities, Inc. The Arecibo Observatory is operated by SRI International under a cooperative agreement with the National Science Foundation (AST-1100968), and in alliance with Ana G. M\'{e}ndez-Universidad Metropolitana, and the Universities Space Research Association. The Nan\c{c}ay Radio Observatory is operated by the Paris Observatory, associated with the French Centre National de la Recherche Scientifique (CNRS). The Effelsberg 100-m telescope is operated by the MPIfR (Max-Planck-Institut f\"ur Radioastronomie). We acknowledge support of GMRT operators. The GMRT is run by the National Centre for Radio Astrophysics of the Tata Institute of Fundamental Research.

\bibliographystyle{apj}
\bibliography{eah_fgrp4_paper2}

\appendix
\input{source_list}

\end{document}

%% file: reloc.tex
\begin{deluxetable*}{llcccccc}[p]
	\tablecaption{\label{t:reloc} Relocalization results}
	\tablecolumns{7}
	\tablehead{
		\colhead{PSR} &
		\colhead{3FGL Source}&
		\colhead{$r_{95}$}&
		\colhead{$r_{95}$}&
		\colhead{$\Delta_\textrm{3FGL}$}&
		\colhead{$\Delta_\textrm{new}$}\\
		\colhead{ }&
		\colhead{ }&
		\colhead{(3FGL)}&
		\colhead{(new)}&
		\colhead{ }&
		\colhead{ }
	}
	\startdata
	J0002+6216   & J0002.6+6218 	&3.6$'$	 & 2.0$'$	&	2.7$'$ 	&	1.3$'$ 	\\
	J0359+5414   & J0359.5+5413	    &2.4$'$	 & 1.8$'$	&	1.8$'$ 	&	0.6$'$ 	\\
	J0631+0646   & J0631.6+0644	    &2.8$'$	 & 1.8$'$	&	4.1$'$ 	&	1.6$'$ 	\\
	J1057$-$5851 & J1056.7$-$5853	&5.2$'$	 & 2.5$'$	&	4.4$'$ 	&	2.9$'$  	\\
	J1105$-$6037 & J1104.9$-$6036	&2.7$'$	 & 1.5$'$	&	0.7$'$ 	&	0.8$'$  	\\
	J1350$-$6225 & J1350.4$-$6224	&2.4$'$	 & 1.6$'$	&	2.0$'$ 	&	2.3$'$ 	\\
	J1528$-$5838 & J1528.3$-$5836	&3.3$'$  & 1.7$'$	&	1.7$'$ 	&	0.1$'$ 	\\
	J1623$-$5005 & J1622.9$-$5004	&1.5$'$	 & 1.0$'$	&	1.7$'$ 	&	1.5$'$ 	\\
	J1624$-$4041 & J1624.2$-$4041	&2.7$'$	 & 1.6$'$	&	0.9$'$ 	&	1.4$'$ 	\\
	J1650$-$4601 & J1650.3$-$4600	&2.1$'$	 & 1.3$'$	&	1.0$'$ 	&	0.9$'$ 	\\
	J1827$-$1446 & J1827.3$-$1446	&3.7$'$	 & 1.6$'$	&	1.2$'$ 	&	0.7$'$ 	\\
	J1844$-$0346 & J1844.3$-$0344	&3.4$'$	 & 1.6$'$	&	2.8$'$	&	2.3$'$	\\
	J2017+3625   & J2017.9+3627	    &2.1$'$	 & 1.2$'$	&	2.4$'$ 	&	0.9$'$ 	\\
	\enddata

	\tablecomments{Results of the relocalization analysis discussed in Section~\ref{s:reloc}. For each of the 13 new pulsars reported in Paper I, column 2 lists the name of the 3FGL source in which the pulsar was discovered. Columns 3 and 4 list the semi-major axis of the 3FGL source error ellipse at 95\% confidence ($r_{95}$) and the semi-major axis value from our analysis. The $r_{95}$ (new) values are based on statistical uncertainties only. Columns 5 and 6 list the offset between pulsar timing positions and 3FGL positions ($\Delta_\textrm{3FGL}$) and the offset between pulsar timing positions and new positions ($\Delta_\textrm{new}$).}

\end{deluxetable*}

%% file: on_pulse.tex
\begin{deluxetable*}{lrrccccl}[p]
	\tablecaption{\label{t:params} On-pulse spectral parameters}
	\tablecolumns{7}
	\tablehead{
		\colhead{PSR}&
		\colhead{TS}&
		\colhead{$\textrm{TS}_\textrm{cut}$}&
		\colhead{$\Gamma$}&
		\colhead{$E_\textrm{cut}$}&
		\colhead{Photon Flux, $F_{100}$}&
		\colhead{Energy Flux, $G_{100}$}\\
		\colhead{ }&
		\colhead{ }&
		\colhead{ }&		
		\colhead{}&
		\colhead{(GeV)}&
		\colhead{($10^{-8}\ \textrm{ph} \textrm{ cm}^{-2} \textrm s^{-1}$)}&
		\colhead{($10^{-11}\ \textrm{erg} \textrm{ cm}^{-2} \textrm s^{-1}$)}
	}
	\startdata

J0002$+$6216&975 & 145&$1.04\,\pm\,0.16\,\pm\,0.14$&$1.39\,\pm\,0.21\,\pm\,0.07$&$2.8\,\pm\,0.3\,\pm\,0.8$&$2.6\,\pm\,0.2\,\pm\,0.4$&\\
J0359$+$5414&1610 & 93&$1.80\,\pm\,0.07\,\pm\,0.10$&$3.72\,\pm\,0.61\,\pm\,0.26$&$8.4\,\pm\,0.6\,\pm\,2.0$&$5.6\,\pm\,0.2\,\pm\,0.8$&\\
J0631$+$0646& 881& 81&$1.30\,\pm\,0.17\,\pm\,0.12$&$3.93\,\pm\,0.84\,\pm\,0.33$&$2.9\,\pm\,0.6\,\pm\,0.6$&$3.7\,\pm\,0.3\,\pm\,0.3$&\\
J1057$-$5851& 813& 123&$1.39\,\pm\,0.16\,\pm\,0.05$&$1.13\,\pm\,0.19\,\pm\,0.09$&$7.9\,\pm\,0.9\,\pm\,0.8$&$5.0\,\pm\,0.3\,\pm\,0.5$&\\
J1105$-$6037&1084 & 94&$1.66\,\pm\,0.11\,\pm\,0.04$&$3.49\,\pm\,0.60\,\pm\,0.26$&$8.3\,\pm\,1.2\,\pm\,0.4$&$6.4\,\pm\,0.5\,\pm\,0.4$&\\
J1350$-$6225& 704& 85&$1.21\,\pm\,0.16\,\pm\,0.44$&$3.80\,\pm\,0.70\,\pm\,1.13$&$4.2\,\pm\,0.8\,\pm\,4.1$&$6.0\,\pm\,0.4\,\pm\,2.1$&\\
J1528$-$5838& 593& 87&$0.97\,\pm\,0.07\,\pm\,0.36$&$2.27\,\pm\,0.12\,\pm\,0.43$&$2.2\,\pm\,0.5\,\pm\,1.3$&$3.0\,\pm\,0.3\,\pm\,0.7$&\\
J1623$-$5005& 854& 106&$1.33\,\pm\,0.01\,\pm\,0.29$&$7.17\,\pm\,0.17\,\pm\,1.48$&$4.7\,\pm\,0.1\,\pm\,3.3$&$8.1\,\pm\,0.2\,\pm\,2.0$&\\
J1624$-$4041& 255& 31&$1.50\,\pm\,0.21\,\pm\,0.38$&$3.59\,\pm\,1.07\,\pm\,0.85$&$1.6\,\pm\,0.8\,\pm\,0.9$&$1.6\,\pm\,0.5\,\pm\,0.5$&\\
J1650$-$4601&1368 &83&$1.70\,\pm\,0.10\,\pm\,0.07$&$4.04\,\pm\,0.71\,\pm\,0.59$&$15.9\,\pm\,1.4\,\pm\,4.6$&$12.3\,\pm\,0.6\,\pm\,2.3$&\\
J1827$-$1446&818 & 134&$0.47\,\pm\,0.28\,\pm\,0.32$&$1.36\,\pm\,0.22\,\pm\,0.15$&$3.7\,\pm\,0.7\,\pm\,1.0$&$5.8\,\pm\,0.5\,\pm\,0.6$&\\
J1844$-$0346&840 & 75&$1.21\,\pm\,0.22\,\pm\,0.23$&$2.59\,\pm\,0.53\,\pm\,0.41$&$8.3\,\pm\,1.9\,\pm\,2.6$&$9.5\,\pm\,0.9\,\pm\,1.5$&\\
J2017+3625&1148 & 216&$0.78\,\pm\,0.15\,\pm\,0.22$&$1.61\,\pm\,0.18\,\pm\,0.18$&$4.7\,\pm\,1.3\,\pm\,1.3$&$6.2\,\pm\,1.1\,\pm\,1.2$&\\

	\enddata

\tablecomments{Binned maximum likelihood spectral fit results for the 13 Einstein@Home gamma-ray pulsars. For each pulsar, columns 2 and 3 list the TS of the source, and the cutoff TS for the exponentially cut-off model over a simple power-law model. Columns 4 and 5 list the best-fit photon index $\Gamma$ and cutoff energy $E_\textrm{cut}$. The next two columns give the on-pulse phase-averaged integral photon and energy fluxes in the 0.1 to 100 GeV band, $F_{100}$ and $G_{100}$, scaled to full interval values. The first reported uncertainties are statistical, while the second uncertainties are systematic, determined by re-analyzing the data with bracketing IRFs and artificially changing the normalization of the Galactic diffuse model by $\pm$6\%, as described in \citet{AceroSNRpaper}.}

\end{deluxetable*}

%% file: off_pulse.tex
\begin{deluxetable*}{lrrcc}[p]
	\tablecaption{\label{t:offpulsefit} Off-pulse spectral parameters}
	\tablecolumns{5}
	\tablehead{
		\colhead{PSR} &
		\colhead{TS}&
		\colhead{$\textrm{TS}_\textrm{cut}$}&
		\colhead{$\Gamma$}&
		\colhead{$E_\textrm{cut}$}\\
		\colhead{}&
		\colhead{}&
		\colhead{}&	
		\colhead{}&
		\colhead{(GeV)}
	}
	\startdata

	J1623$-$5005& 	57	& 18& *							 &$0.87\, \pm\, 0.07\, \pm\, 0.21$\\
	J1624$-$4041& 47	& 10&$1.02\,\pm\,0.95\,\pm\,0.96$&$1.33\,\pm\,1.23\,\pm\,0.41$\\
	J2017$+$3625& 215	& 88&$0.69\,\pm\,0.06\,\pm\,0.06$&$0.59\,\pm\,0.01\,\pm\,0.06$\\
	\enddata

	\tablecomments{Results of the maximum likelihood analysis of the off-pulse phase ranges of pulsars with significant off-pulse emission, as discussed in Section~\ref{s:spectral_analysis}. Column 1 lists the name of the pulsar. Columns 2-5 list the TS of the source in the off-pulse phase range, the test statistic $\textrm{TS}_\textrm{cut}$ of an exponentially cut-off model over a simple power-law model, the photon index $\Gamma$ and energy cutoff $E_\mathrm{cut}$.\\
	$*$ Although the spectral index is consistent with zero, the well-defined $E_\mathrm{cut}$ allows integration to a finite total flux.}

\end{deluxetable*}

%% file: pulse_shape.tex
\begin{deluxetable*}{lcccccccc}[p]

	\tablecaption{\label{t:pulse_shape_parameters} Pulse shape parameters and Derived Pulsar Parameters}
	\tablecolumns{8}
	\tablehead{
		\colhead{PSR}&
		\colhead{Peaks}&
		\colhead{$\delta$}&
		\colhead{$\Delta$}&
		\colhead{Off-pulse phase range}&
		\colhead{$\dot E$}&
		\colhead{DM distance}&
		\colhead{Heuristic distance, $d_h$}\\
		\colhead{}&
		\colhead{}&
		\colhead{}&
		\colhead{}&
		\colhead{}&
		\colhead{$(10^{33} \textrm{erg} ~\textrm s^{-1})$}&
		\colhead{(kpc)}&
		\colhead{(kpc)}
}
\startdata
	J0002+6216   &2	&0.171 $\pm$ 0.011	&0.361 $\pm$ 0.012&0.59--1.00				&153	&7.7, 6.3	&2.0\\
	J0359+5414   &1	&...				&...			  &0.00--0.58				&1318	&...		&2.3\\
	J0631+0646   &2	&0.469 $\pm$ 0.013	&0.278 $\pm$ 0.013&0.83--0.31				&104	&>42.2, 4.6	&1.5\\
	J1057$-$5851 &1	&...				&...			  &0.75--0.24				&17		&...		&0.8\\
	J1105$-$6037 &2	&...				&0.317 $\pm$ 0.006&0.90--0.38				&116	&...		&1.2\\
	J1350$-$6225 &2	&...				&0.485 $\pm$ 0.002&0.92--0.24, 0.52--0.77 	&133	&...		&1.3\\
	J1528$-$5838 &2	&...				&0.243 $\pm$ 0.022&0.48--1.00				&22		&...		&1.1\\
	J1623$-$5005 &2	&...				&0.352 $\pm$ 0.005&0.99--0.45				&267	&...		&1.3\\
	J1624$-$4041 &2	&...				&0.429 $\pm$ 0.003&0.44--0.70				&39		&...		&1.8\\
	J1650$-$4601 &2	&...				&0.331 $\pm$ 0.005&0.48--1.00				&291	&...		&1.1\\
	J1827$-$1446 &2	&...				&0.256 $\pm$ 0.008&0.82--0.32				&14 	&...		&0.7\\
	J1844$-$0346 &1	&...				&...			  &0.31--0.92				&4249	&...		&2.4\\
	J2017+3625   &2	&...				&0.374 $\pm$ 0.004&0.02--0.42, 0.58--0.68 	&12		&...		&0.7\\
	\enddata
	\tablecomments{Columns 2-5 list the gamma-ray peak multiplicity, radio-to-gamma-ray phase lag ($\delta$), gamma-ray peak separation ($\Delta$) for pulse profiles with two components, and definition of the off-pulse phase interval, for each pulsar considered in our study. Uncertainties on $\delta$ and $\Delta$ are statistical only. Column 6 gives the spin-down power for each pulsar. Column 7 lists the DM distances for the radio-detected pulsars J0002+6216 and J0631+1036 as inferred with the NE2001 model of \citet{Cordes2002} and the model of \citet{ymw17}. The last column lists the heuristic distance, described in Sec. \ref{s:luminosity}.}
		
\end{deluxetable*}

%% file: radio_followup_def.tex
\begin{deluxetable*}{llccccccc}[p]
\tablecaption{\label{t:radio_follow_def} Definition of Radio Observation Codes}
\tablecolumns{9}
\tablehead{
	\colhead{Obs Code} & 
	\colhead{Telescope} & 
	\colhead{Gain} & 
	\colhead{Frequency} & 
	\colhead{Bandwidth $\Delta F$} & 
	\colhead{$\beta$} & 
	\colhead{$n_p$} & 
	\colhead{HWHM} & 
	\colhead{$T_\mathrm{rec}$}\\ 
	\colhead{} & 
	\colhead{} & 
	\colhead{(K Jy$^{-1}$)} & 
	\colhead{(MHz)} & 
	\colhead{(MHz)} & 
	\colhead{} & 
	\colhead{} & 
	\colhead{(arcmin)} & 
	\colhead{(K)}
}
\startdata
AO-327 & Arecibo & $11$ & $327$ & $68$ & $1.12$ & $2$ & $6.3$ & $116$ \\
AO-ALFA & Arecibo & $10$ & $1400$ & $100$ & $1.12$ & $2$ & $1.5$ & $30$ \\
AO-Lwide & Arecibo & $10$ & $1510$ & $300$ & $1.12$ & $2$ & $1.5$ & $27$ \\
Eff-7B & Effelsberg & $1.5$ & $1400$ & $240$ & $1.05$ & $2$ & $9.1$ & $22$ \\
Eff-L1 & Effelsberg & $1.5$ & $1400$ & $240$ & $1.05$ & $2$ & $9.1$ & $22$ \\
GBT-820 & GBT & $2.0$ & $820$ & $200$ & $1.05$ & $2$ & $7.9$ & $29$ \\
GBT-S & GBT & $1.9$ & $2000$ & $700$ & $1.05$ & $2$ & $3.1$ & $22$ \\
GMRT-322 & GMRT & $1.6$ & $322$ & $32$ & $1$ & $2$ & $40$ & $106$ \\
GMRT-610 & GMRT & $1.6$ & $607$ & $32$ & $1$ & $2$ & $20$ & $102$ \\
Nancay-L & Nancay & $1.4$ & $1398$ & $128$ & $1.05$ & $2$ & $2 \times 11$ & $35$ \\
Parkes-AFB & Parkes & $0.735$ & $1374$ & $288$ & $1.25$ & $2$ & $7$ & $25$ \\
Parkes-BPSR & Parkes & $0.735$ & $1352$ & $340$ & $1.05$ & $2$ & $7$ & $25$ \\
Parkes-DFB4 & Parkes & $0.735$ & $1369$ & $256$ & $1.1$ & $2$ & $7$ & $25$ \\
\enddata
\tablecomments{Radio telescopes and backend parameters used for follow-up observations of the new pulsars, described in Section~\ref{s:radio}.}
\end{deluxetable*}

%% file: radio_followup_v3.tex
\begin{deluxetable*}{lllcccccc}[p]
\tablecaption{\label{t:radio_follow_up} Radio Search Observations of the New Pulsars}
\tablecolumns{9}
\tablehead{
	\colhead{Target} & 
	\colhead{Obs Code} & 
	\colhead{Date} & 
	\colhead{$t_\mathrm{int}$} & 
	\colhead{R.A.} & 
	\colhead{Decl.} & 
	\colhead{Offset} & 
	\colhead{$T_\mathrm{sky}$} & 
	\colhead{$S_\mathrm{min}$}\\ 
	\colhead{PSR} & 
	\colhead{} & 
	\colhead{} & 
	\colhead{(min)} & 
	\colhead{(J2000)} & 
	\colhead{(J2000)} & 
	\colhead{(arcmin)} & 
	\colhead{(K)} & 
	\colhead{($\mu$Jy)}
}
\startdata
J0002+6216 & GBT-S & 2013 Feb 28 & $28$ & 00:02:40.3 & 62:16:44.0 & $2.2$ & $0.9$ & Detected \\
 & Eff-L1 & 2015 Feb 14 & $120$ & 00:02:58.1 & 62:16:09.6 & $0.0$ & $2.4$ & Detected \\
\hline
J0359+5414 & Eff-7B & 2010 May 25 & $32$ & 03:59:35.8 & 54:10:40.8 & $4.5$ & $2.0$ & $34$ \\
 & Eff-7B & 2010 Jul 17 & $60$ & 03:59:31.5 & 54:11:44.1 & $3.3$ & $2.0$ & $23$ \\
 & GBT-S & 2012 Nov 17 & $40$ & 03:59:36.3 & 54:12:56.5 & $2.5$ & $0.8$ & $18$ \\
 & GBT-S & 2013 Mar 17 & $7$ & 03:59:36.3 & 54:12:56.5 & $2.5$ & $0.8$ & $42$ \\
 & Eff-L1 & 2015 Feb 14 & $115$ & 03:59:26.0 & 54:14:55.6 & $0.0$ & $2.0$ & $15$ \\
\hline
J0631+0646 & Eff-L1 & 2015 Feb 13 & $120$ & 06:31:52.4 & 06:46:15.3 & $0.0$ & $1.8$ & Detected \\
 & AO-327 & 2015 Mar 15 & $75$ & 06:31:52.4 & 06:46:14.0 & $0.0$ & $78.4$ & Detected \\
 & AO-Lwide & 2015 Jun 14 & $69$ & 06:31:52.4 & 06:46:14.0 & $0.0$ & $1.5$ & Detected \\
\hline
J1057$-$5851 & Parkes-DFB4 & 2015 Aug 05 & $70$ & 10:57:09.3 & $-$58:51:11.0 & $0.1$ & $3.9$ & $49$ \\
 & Parkes-DFB4 & 2015 Aug 06 & $51$ & 10:57:09.3 & $-$58:51:11.0 & $0.1$ & $3.9$ & $58$ \\
\hline
J1105$-$6037 & Parkes-DFB4 & 2015 Aug 05 & $70$ & 11:05:00.5 & $-$60:37:15.6 & $0.0$ & $5.7$ & $52$ \\
 & Parkes-DFB4 & 2015 Aug 06 & $60$ & 11:05:00.5 & $-$60:37:15.6 & $0.0$ & $5.7$ & $56$ \\
\hline
J1350$-$6225 & Parkes-AFB & 2010 Nov 19 & $145$ & 13:49:36.0 & $-$62:24:00.0 & $8.1$ & $10.5$ & $110$ \\
 & Parkes-BPSR & 2010 Nov 19 & $144$ & 13:49:36.0 & $-$62:24:00.0 & $8.1$ & $10.9$ & $86$ \\
 & Parkes-DFB4 & 2015 Sep 05 & $433$ & 13:50:44.5 & $-$62:25:43.7 & $0.0$ & $10.6$ & $24$ \\
 & Parkes-DFB4 & 2015 Sep 13 & $500$ & 13:50:44.5 & $-$62:25:43.7 & $0.0$ & $10.6$ & $23$ \\
\hline
J1623$-$5005 & Parkes-BPSR & 2010 Nov 19 & $144$ & 16:22:48.0 & $-$50:06:00.0 & $2.7$ & $16.9$ & $45$ \\
 & Parkes-AFB & 2010 Nov 19 & $88$ & 16:22:48.0 & $-$50:06:00.0 & $2.7$ & $16.2$ & $74$ \\
\hline
J1624$-$4041 & Parkes-AFB & 2009 Dec 02 & $120$ & 16:24:06.2 & $-$40:40:48.0 & $1.0$ & $4.1$ & $41$ \\
 & GBT-S & 2009 Dec 23 & $30$ & 16:24:06.0 & $-$40:40:48.0 & $1.0$ & $1.5$ & $15$ \\
 & Parkes-AFB & 2010 Jul 18 & $120$ & 16:24:03.0 & $-$40:42:56.0 & $1.9$ & $4.1$ & $43$ \\
 & Parkes-AFB & 2010 Jul 26 & $120$ & 16:24:03.0 & $-$40:42:56.0 & $1.9$ & $4.1$ & $43$ \\
 & Parkes-AFB & 2010 Nov 12 & $60$ & 16:24:03.0 & $-$40:42:56.0 & $1.9$ & $4.1$ & $60$ \\
 & GMRT-610 & 2011 Feb 15 & $60$ & 16:24:03.8 & $-$40:41:20.4 & $1.2$ & $34.4$ & $297$ \\
 & Parkes-AFB & 2012 Jul 12 & $60$ & 16:24:09.0 & $-$40:40:23.0 & $1.1$ & $4.1$ & $58$ \\
 & GMRT-322 & 2012 Jul 12 & $60$ & 16:24:09.0 & $-$40:40:23.0 & $1.1$ & $178.6$ & $618$ \\
 & Parkes-AFB & 2012 Dec 17 & $60$ & 16:24:09.0 & $-$40:40:23.0 & $1.1$ & $4.1$ & $58$ \\
\hline
J1650$-$4601 & Parkes-BPSR & 2010 Nov 21 & $144$ & 16:50:48.0 & $-$46:06:00.0 & $6.9$ & $14.4$ & $74$ \\
 & Parkes-AFB & 2010 Nov 21 & $139$ & 16:50:48.0 & $-$46:06:00.0 & $6.9$ & $13.8$ & $96$ \\
\hline
J1827$-$1446 & GBT-820 & 2014 Apr 21 & $35$ & 18:27:20.2 & $-$14:46:01.2 & $1.2$ & $33.2$ & $60$ \\
 & Eff-L1 & 2015 Feb 14 & $120$ & 18:27:24.6 & $-$14:46:25.4 & $0.0$ & $8.3$ & $19$ \\
\hline
J1844$-$0346 & Eff-7B & 2010 May 15 & $32$ & 18:44:15.4 & $-$03:42:46.8 & $5.7$ & $11.8$ & $53$ \\
 & Eff-7B & 2010 Jul 30 & $60$ & 18:44:21.8 & $-$03:42:03.6 & $5.2$ & $11.8$ & $37$ \\
 & Eff-7B & 2010 Jul 31 & $60$ & 18:44:21.8 & $-$03:42:03.6 & $5.2$ & $11.8$ & $37$ \\
 & GBT-S & 2012 Nov 17 & $22$ & 18:44:26.2 & $-$03:45:21.6 & $2.0$ & $4.7$ & $24$ \\
 & Eff-7B & 2015 Aug 27 & $120$ & 18:44:33.0 & $-$03:46:32.0 & $0.0$ & $11.8$ & $21$ \\
\hline
J2017+3625 & Nancay-L & 2010 May 05 & $65$ & 20:17:55.8 & 36:25:08.0 & $0.0$ & $4.6$ & $50$ \\
 & Nancay-L & 2010 May 11 & $47$ & 20:17:55.8 & 36:25:08.0 & $0.0$ & $4.6$ & $58$ \\
 & GBT-S & 2010 May 13 & $60$ & 20:17:59.0 & 36:25:19.0 & $0.7$ & $1.8$ & $10$ \\
 & GBT-820 & 2011 Jan 15 & $45$ & 20:17:57.6 & 36:27:36.0 & $2.5$ & $18.6$ & $43$ \\
 & AO-ALFA & 2015 May 11 & $20$ & 20:17:54.2 & 36:23:24.0 & $1.8$ & $4.6$ & $34$ \\
 & AO-327 & 2015 Jun 24 & $15$ & 20:17:55.9 & 36:27:32.4 & $2.4$ & $202.7$ & $170$ \\
 & AO-327 & 2015 Jun 25 & $15$ & 20:17:55.9 & 36:27:32.4 & $2.4$ & $202.7$ & $170$ \\
 & Eff-7B & 2015 Aug 27 & $120$ & 20:17:55.8 & 36:25:08.0 & $0.0$ & $4.6$ & $17$ \\
 & AO-327 & 2015 Nov 16 & $28$ & 20:17:55.9 & 36:25:08.4 & $0.0$ & $202.7$ & $113$ \\
 & AO-Lwide & 2015 Nov 17 & $33$ & 20:17:55.9 & 36:25:08.4 & $0.0$ & $3.8$ & $5$ \\
\enddata
\tablecomments{Radio observations of the new pulsars. In the cases of PSRs~J0002+6216 and J0631+0646, radio pulsations were detected (see Section~\ref{s:radio}).}
\end{deluxetable*}

%% file: profile_fitting.tex
\begin{deluxetable*}{l c c c c c c c c c c}
	\tablewidth{0pt}
	\tablecaption{Light Curve Modeling Results \label{tab:models}}
	\tablecolumns{11}
	\tablehead{\colhead{PSR} & \colhead{TPC $-\ln\mathcal{L}$} & \colhead{TPC $\alpha$} & \colhead{TPC $\zeta$} & \colhead{TPC $f_{\Omega}$} & \colhead{TPC Radio Flag} & \colhead{OG $-\ln\mathcal{L}$} & \colhead{OG $\alpha$} & \colhead{OG $\zeta$} & \colhead{OG $f_{\Omega}$} & \colhead{OG Radio Flag}\\ \colhead{} & \colhead{} & \colhead{($^\circ$)} & \colhead{($^\circ$)} & \colhead{} & \colhead{} & \colhead{} & \colhead{($^\circ$)} & \colhead{($^\circ$)} & \colhead{} & \colhead{}}
	\startdata
	J0002+6216 & 110.26 & $64^{+3}_{-2}$ & $54\pm2$ & $1.05\pm0.04$ & \nodata & 105.70 & $69^{+8}_{-1}$ & $58^{+25}_{-1}$ & $1.08^{+0.05}_{-0.27}$ & \nodata \\
	J0359+5414$^{\dag}$ & 39.88 & $1\pm1$ & $2\pm1$ & $19.62^{+0.01}_{-8.52}$ & L & 38.04 & $80^{+8}_{-6}$ & $24\pm4$ & $1,01^{+0.09}_{-0.41}$ & Q \\
	J1057$-$5851$^{\dag}$ & 32.62 & $57^{+2}_{-3}$ & $40^{+7}_{-2}$ & $0.95^{+0.05}_{-0.18}$ & F & 42.94 & $65^{+2}+{-1}$ & $28^{+1}_{-2}$ & $0.76^{+0.10}_{-0.03}$ & Q \\
	J1105$-$6037 & 46.11 & $61^{+4}_{-27}$ & $49^{+21}_{-7}$ & $0.98^{+0.05}_{-0.31}$ & F & 67.40 & $8^{+5}_{-2}$ & $71^{+4}_{-1}$ & $0.99^{+0.01}_{-0.09}$ & Q \\
	J1350$-$6225 & 79.42 & $82^{+2}_{-4}$ & $85^{+1}_{-2}$ & $0.82\pm0.10$ & L & 48.16 & $90\pm9$ & $88^{+1}_{4}$ & $0.70\pm0.03$ & L \\
	J1528$-$5838$^{\dag}$ & 29.71 & $2\pm1$ & $2\pm1$ & $3.77^{+0.01}_{-0.28}$ & L & 27.21 & $9^{+9}_{-6}$ & $74^{+6}_{-3}$ & $0.95^{+0.04}_{-0.09}$ & Q \\
	J1623$-$5005 & 31.28 & $32^{+2}_{-1}$ & $68\pm1$ & $0.62^{+0.02}_{-0.01}$ & Q & 58.83 & $9^{+12}_{-1}$ & $72^{+3}_{-1}$ & $0.21^{+0.19}_{-0.01}$ & Q \\
	J1624$-$4041 & 86.57 & $71^{+2}_{-5}$ & $58^{+1}_{-5}$ & $1.13\pm0.03$ & F & 72.90 & $86\pm1$ & $68\pm1$ & $1.02^{+0.02}_{-0.01}$ & F \\
	J1650$-$4601 & 46.30 & $13^{+2}_{-7}$ & $69\pm1$ & $0.47^{+0.01}_{-0.09}$ & Q & 54.13 & $11^{+2}_{-4}$ & $74^{+6}_{-3}$ & $0.21^{+0.19}_{-0.16}$ & Q \\
	J1827$-$1446$^{\dag}$ & 52.65 & $1\pm1$ & $2\pm1$ & $69.16^{+0.01}_{-5.67}$ & L & 45.04 & $75^{+1}_{-11}$ & $26^{+5}_{-1}$ & $1.34^{+0.01}_{-0.71}$ & Q \\
	J1844$-$0346$^{\dag}$ & 23.06 & $10\pm1$ & $68\pm1$ & $0.49\pm0.07$ & Q & 22.08 & $79^{+6}_{-4}$ & $22^{+1}_{-3}$ & $0.99^{+0.31}_{-0.39}$ & Q \\
	J2017+3625 & 168.10 & $23\pm5$ & $69\pm1$ & $0.52^{+0.16}_{-0.01}$ & Q & 127.47 & $16^{+12}_{-5}$ & $80^{+1}_{-5}$ & $0.23^{+0.10}_{-0.04}$ & Q \\
		\enddata
		\tablecomments{Light curve fitting results for all pulsars except PSR J0631+0646. Column 1 gives the pulsar name, a $\dag$ indicates that the $\Delta\ln\mathcal{L}$ surface was not renormalized.  Column 2 (7) gives the best-fit $-\ln\mathcal{L}$ value for the TPC (OG) model.  Columns 3, 4, and 5 (8, 9, and 10) give the best-fit $\alpha$ and $\zeta$ with corresponding $f_{\Omega}$ for the TPC (OG) model.  For pulsars without a radio detection, column 6 (11) gives a radio-loudness prediction from the best-fit geometry for the TPC (OG) model: L = radio-loud, F = radio-faint, and Q = radio-quiet; see the text for details.}
\end{deluxetable*}

%% file: x-ray.tex
\begin{deluxetable*}{ccccccc}[p]
	\tablewidth{0pt}
	\tablecaption{Summary of the pulsar X-ray spectral parameters}
	\tablecolumns{7}
	\tablehead{
		\colhead{\multirow{2}{*}{PSR}} & 
		\colhead{X-ray} & 
		\colhead{Exposure} & 
		\colhead{N$_\textrm{H}$} & 
		\colhead{\multirow{2}{*}{$\Gamma_{X}$}} & 
		\colhead{F$_X$$^b$} & 
		\colhead{\multirow{2}{*}{G$_{100}$/F$_X$$^c$}}\\
		\colhead{}&
		\colhead{observatory$^a$}&
		\colhead{(ksec)}&
		\colhead{($10^{21}$ cm$^{-2}$)}&
		\colhead{}&
		\colhead{($10^{-14}$\,erg cm$^{-2}$ s$^{-1}$)}&
		\colhead{}
		}
	\startdata
	
	J0002+6216 & \emph{Swift} & 9.2 & 1.0$^c$ & 2$^c$ & 4.3$^{+1.9}_{-2.4}$ 		&  600$^{+760}_{-200}$	\\
	J0359+5414 & \emph{Chandra} & 29.7 & 2.0$\pm$1.1 & 2.7$\pm$0.3 & 0.96$\pm0.20$  & 5800$\pm$1500			\\
	J0631+0646 & \emph{Swift} & 3.5 & 0.4$^c$ & 2$^c$ & $<$9.1 						& $>$400 				\\
	J1057--5851 & \emph{Chandra} & 10.1 & 3.0$^c$ & 2$^c$ & $<$0.25 				& $>$20000 				\\
	J1105--6037 & \emph{Swift} & 16 & 1.0$^c$ & 2$^c$ & 4.8$^{+1.9}_{-1.4}$ 		&  1300$^{+550}_{-380}$	\\
	J1350--6225 & \emph{Swift} & 5.4 & 1.4$^c$ & 2$^c$ & $<$8.1 					& $>$740				\\
	J1528--5838 & \emph{Swift} & 6 & 0.7$^c$ & 2$^c$ & $<$6.2 						& $>$480				\\
	J1623--5005 & \emph{XMM-Newton} & 85.4 & 4.0$^c$ & 2$^c$ 	& $<$2.0 			& $>$4100				\\
	\multirow{2}{*}{J1624--4041} & \multirow{2}{*}{\emph{XMM-Newton}} & \multirow{2}{*}{31.0} & \multirow{2}{*}{2.0$^c$} & (src1)  0.7$\pm0.2$ & 3.7$\pm$0.7 & 430$\pm$100\\
	& & & & (src2)  2.0$\pm0.4$ & 1.0$\pm0.3$ & 1600$\pm$690\\
	J1650--4601 & \emph{Swift} & 3.5 & 1.0$^c$ & 2$^d$ & $<$10.8 & $>$1100\\
	J1827--1446 & -- & -- & -- & -- & -- & --\\
	J1844--0344 & \emph{Swift} & 82 & 2.4$^c$ & 2$^c$ & 7.6$\pm$1.3 & 1300$\pm$260 \\
	J2017+3625 & \emph{Chandra} & 10.0 & 1.0$^c$ & 2$^c$ & 1.7$\pm0.7$ & 3600$\pm$2600 \\
	
	\enddata
	\tablecomments{Results of the analysis of archival X-ray observations. Columns 1 and 2 list the pulsar name and the X-ray observatory. Columns 3-5 list the duration of the exposure, and for each X-ray counterpart the best-fit column density and photon index. The following two columns give the unabsorbed X-ray flux in the 0.3--10 keV energy band, and the gamma-ray-to-X-ray flux ratio. All uncertainties are reported at the 68\% confidence level.\vspace{.2cm}\\
	$^a$We report only the X-ray observatory used for the spectral analysis.\\
	$^b$When the X-ray counterpart is not detected, we report the minimum
	X-ray unabsorbed flux required for a 3$\sigma$ detection.\\
	$^c$Gamma-ray energy fluxes in 0.1--100 GeV are used to calculate the gamma-ray-to-X-ray flux ratio. \\
	$^d$Due to the low statistics in these sources, we fixed this parameter in the spectral analysis as described in the text.\label{tab:1}}

\end{deluxetable*}

%% file: source_list.tex

\LongTables

\begin{deluxetable*}{lcccrrrrrrcc}[p]		
	
	\tablecaption{\label{t:source_list} Ranked source list}
	\tablecolumns{10}
	\tablehead{
		\colhead{3FGL Name}&
		\colhead{Searched R.A.}&
		\colhead{Searched Decl.}&
		\colhead{Search Radius}&
		\colhead{VI}&		
		\colhead{$\textrm{TS}_\textrm{curve}$}&
		\colhead{$\textrm{TS}_\textrm{cut}$}&
		\colhead{$E_\textrm{cut}$}&
		\colhead{$\Gamma$}&
		\colhead{TS}&
		\colhead{log $R_{S}$}&
		\colhead{Class}\\
		\colhead{}&
		\colhead{(J2000)}&
		\colhead{(J2000)}&
		\colhead{(arcmin)}&
		\colhead{}&
		\colhead{}&
		\colhead{}&
		\colhead{(GeV)}&
		\colhead{}&
		\colhead{}&
		\colhead{}&
		\colhead{}
	}

J1745.3$-$2903c						&	17:45:22.32	&	$-$29:03:46.80	&	2.05	&	48.42	&	275.2	&	378.7	&	2.2	&	1.4	&	3407&	18.85	&	...	\\
J1746.3$-$2851c						&	17:46:22.51	&	$-$28:51:45.72	&	2.12	&	57.06	&	113.0	&	364.7	&	4.0	&	1.5	&	2373	&	14.31	&	pwn	\\
\textbf{J2017.9$+$3627}				&	20:17:56.33	&	$+$36:27:32.76	&	3.10	&	39.86	&	179.3	&	198.9	&	1.9	&	1.4	&	1876	&	13.61	&	...	\\
J1839.3$-$0552$^\dagger$			&	18:39:23.52	&	$-$05:52:53.76	&	3.07	&	37.43	&	83.7	&	135.1	&	2.3	&	1.2	&	714	&	13.26	&	...	\\
\textbf{J1906.6$+$0720}$^\dagger$	&	19:06:41.14	&	$+$07:20:02.04	&	3.33	&	41.70	&	87.9	&	68.6	&	7.0	&	2.0	&	1580	&	12.51	&	...	\\
J1910.9$+$0906$^\dagger$			&	19:10:58.61	&	$+$09:06:01.80	&	1.55	&	52.13	&	53.2	&	17.4	&	41.7&	2.1	&	4790	&	12.31	&	snr	\\
J1636.2$-$4734$^\dagger$			&	16:36:16.49	&	$-$47:34:49.08	&	4.58	&	54.63	&	106.0	&	47.1	&	7.1	&	1.9	&	1180	&	12.28	&	snr	\\
J1848.4$-$0141						&	18:48:28.39	&	$-$01:41:33.72	&	7.27	&	52.63	&	109.0	&	13.8	&	9.8	&	2.5	&	1457	&	11.81	&	...	\\
J1405.4$-$6119$^\dagger$			&	14:05:25.46	&	$-$61:19:00.48	&	2.83	&	43.93	&	61.1	&	19.7	&	8.2	&	2.1	&	1671	&	11.39	&	...	\\
J1111.9$-$6038$^\dagger$			&	11:11:58.44	&	$-$60:38:27.96	&	1.96	&	46.69	&	81.4	&	58.5	&	10.4&	1.9	&	3624	&	11.36	&	spp	\\
J1748.3$-$2815c						&	17:48:22.20	&	$-$28:15:32.04	&	2.73	&	34.06	&	77.4	&	68.6	&	4.7	&	1.4	&	489	&	11.26	&	...	\\
\textbf{J1622.9$-$5004}$^\dagger$	&	16:22:54.31	&	$-$50:04:31.08	&	2.17	&	54.35	&	72.4	&	73.3	&	8.0	&	1.6	&	891	&	10.21	&	...	\\
J0223.6$+$6204$^\dagger$			&	02:23:37.46	&	$+$62:04:51.96	&	3.51	&	41.77	&	86.3	&	182.6	&	1.8	&	1.5	&	1089	&	9.78	&	...	\\
J1823.2$-$1339$^\dagger$			&	18:23:16.90	&	$-$13:39:04.68	&	2.60	&	47.54	&	29.7	&	47.4	&	9.0	&	1.9	&	1004	&	9.72	&	...	\\
J1745.1$-$3011						&	17:45:11.30	&	$-$30:11:57.84	&	6.17	&	59.68	&	92.7	&	88.8	&	0.6	&	0.4	&	459	&	9.69	&	spp	\\
J1800.8$-$2402$^\dagger$			&	18:00:53.18	&	$-$24:02:06.36	&	3.13	&	46.65	&	36.4	&	21.3	&	11.3&	1.7	&	575	&	9.69	&	...	\\
J1749.2$-$2911						&	17:49:15.58	&	$-$29:11:34.44	&	7.21	&	41.77	&	50.9	&	43.6	&	1.6	&	1.3	&	265	&	9.62	&	...	\\
J1306.4$-$6043$^\dagger$			&	13:06:27.50	&	$-$60:43:54.12	&	2.48	&	35.69	&	65.9	&	42.6	&	8.6	&	1.7	&	1108	&	9.59	&	...	\\
\textbf{J1104.9$-$6036}$^\dagger$	&	11:04:59.42	&	$-$60:36:32.76	&	4.10	&	43.09	&	77.4	&	64.6	&	3.6	&	1.7	&	769	&	9.42	&	...	\\
J0634.1$+$0424						&	06:34:06.79	&	$+$04:24:22.32	&	9.77	&	42.87	&	123.3	&	60.2	&	1.8	&	2.2	&	1421	&	9.41	&	...	\\
J1552.8$-$5330						&	15:52:50.90	&	$-$53:30:47.16	&	6.98	&	46.44	&	56.6	&	50.3	&	1.8	&	1.0	&	210	&	9.26	&	...	\\
J1747.0$-$2828$^\dagger$			&	17:47:05.98	&	$-$28:28:54.84	&	3.65	&	90.61	&	159.7	&	135.3	&	2.5	&	1.8	&	1676	&	9.22	&	...	\\
\textbf{J1650.3$-$4600}				&	16:50:23.76	&	$-$46:00:50.76	&	3.14	&	55.06	&	54.6	&	55.0	&	4.8	&	1.8	&	897	&	9.19	&	...	\\
J2323.4$+$5849						&	23:23:28.85	&	$+$58:49:09.48	&	1.49	&	40.07	&	62.4	&	39.1	&	26.4&	1.6	&	2568	&	9.17	&	snr	\\
J1625.1$-$0021$^\dagger$			&	16:25:07.06	&	$-$00:21:30.96	&	3.38	&	37.31	&	104.3	&	201.4	&	1.9	&	0.8	&	1778	&	8.98	&	...	\\
J1714.5$-$3832						&	17:14:34.27	&	$-$38:32:55.68	&	2.65	&	68.77	&	39.3	&	23.3	&	14.7&	2.2	&	2649	&	8.95	&	snr	\\
J1857.9$+$0210$^\dagger$			&	18:57:57.65	&	$+$02:10:13.44	&	5.41	&	50.62	&	42.8	&	50.5	&	3.2	&	1.9	&	601	&	8.89	&	...	\\
\textbf{J1056.7$-$5853}				&	10:56:42.86	&	$-$58:53:45.60	&	7.77	&	35.71	&	88.2	&	126.1	&	1.1	&	1.5	&	596	&	8.83	&	...	\\
J1026.2$-$5730						&	10:26:14.33	&	$-$57:30:59.76	&	4.85	&	50.42	&	54.7	&	58.1	&	2.3	&	1.6	&	493	&	8.26	&	...	\\
J1742.6$-$3321						&	17:42:39.60	&	$-$33:21:22.32	&	6.00	&	48.24	&	67.1	&	24.4	&	2.5	&	1.8	&	411	&	8.20	&	...	\\
\textbf{J1844.3$-$0344$^\dagger$}	&	18:44:23.93	&	$-$03:44:48.48	&	5.09	&	44.78	&	37.0	&	70.9	&	1.9	&	0.8	&	468	&	8.12	&	...	\\
J1101.9$-$6053						&	11:01:55.46	&	$-$60:53:45.96	&	7.49	&	23.32	&	40.8	&	61.3	&	2.4	&	1.8	&	519	&	7.95	&	spp	\\
J2038.4$+$4212						&	20:38:29.95	&	$+$42:12:30.60	&	5.30	&	45.67	&	51.1	&	95.8	&	0.5	&	0.6	&	340	&	7.92	&	...	\\
J1849.4$-$0057						&	18:49:25.30	&	$-$00:57:06.48	&	3.55	&	45.11	&	23.8	&	16.6	&	13.5&	2.0	&	674	&	7.86	&	snr	\\
J1112.0$-$6135						&	11:12:04.03	&	$-$61:35:03.12	&	8.87	&	55.72	&	84.6	&	35.8	&	1.7	&	1.7	&	293	&	7.84	&	...	\\
J1754.0$-$2538						&	17:54:02.02	&	$-$25:38:54.96	&	2.62	&	66.89	&	72.4	&	107.3	&	4.0	&	1.0	&	500	&	7.73	&	...	\\
J0854.8$-$4503$^\dagger$			&	08:54:50.59	&	$-$45:03:41.76	&	4.37	&	44.94	&	47.5	&	54.9	&	5.0	&	1.7	&	737	&	7.68	&	...	\\
J1857.2$+$0059						&	18:57:14.28	&	$+$00:59:10.68	&	3.82	&	57.14	&	32.6	&	113.2	&	4.5	&	1.3	&	383	&	7.67	&	...	\\
J1740.5$-$2843						&	17:40:30.00	&	$-$28:43:01.20	&	5.87	&	46.42	&	25.6	&	24.2	&	3.6	&	2.2	&	700	&	7.66	&	...	\\
\textbf{J1744.1$-$7619}$^\dagger$	&	17:44:10.85	&	$-$76:19:42.96	&	3.12	&	51.73	&	112.5	&	169.2	&	2.1	&	1.2	&	1759&	7.61	&	...	\\
\textbf{J1035.7$-$6720}$^\dagger$	&	10:35:42.24	&	$-$67:20:00.60	&	3.34	&	47.01	&	80.6	&	120.2	&	2.3	&	1.4	&	1336&	7.39	&	...	\\
J1843.7$-$0322						&	18:43:42.77	&	$-$03:22:37.92	&	7.67	&	70.63	&	65.5	&	54.5	&	3.7	&	2.6	&	1113&	7.37	&	...	\\
\textbf{J0359.5$+$5413}$^\dagger$	&	03:59:31.46	&	$+$54:13:19.20	&	3.66	&	33.63	&	42.2	&	84.1	&	2.6	&	1.6	&	800	&	7.19	&	...	\\
\textbf{J1624.2$-$4041}$^\dagger$	&	16:24:14.26	&	$-$40:41:11.40	&	4.02	&	50.80	&	58.8	&	74.2	&	2.8	&	1.6	&	945	&	7.18	&	...	\\
J1740.5$-$2726						&	17:40:32.28	&	$-$27:27:00.00	&	8.30	&	43.15	&	39.9	&	31.1	&	1.8	&	2.0	&	401	&	7.04	&	...	\\
\textbf{J1827.3$-$1446}				&	18:27:20.16	&	$-$14:46:01.92	&	5.54	&	40.00	&	18.2	&	83.5	&	2.5	&	1.4	&	483	&	6.96	&	...	\\
\hline\\
J2032.5$+$3921						&	20:32:29.78	&	$+$39:25:20.60	&	3.69	&	49.41	&	46.2	&	34.1	&	0.4	&	0.8	&	233	&	6.95	&	...	\\
J1638.6$-$4654						&	16:38:40.16	&	$-$46:54:06.33	&	2.24	&	77.58	&	48.0	&	46.8	&	3.7	&	1.8	&	614	&	6.84	&	spp	\\
J1925.4$+$1727						&	19:24:58.98	&	$+$17:24:41.84	&	7.38	&	47.33	&	42.2	&	22.3	&	1.2	&	1.2	&	157	&	6.70	&	...	\\
J1857.9$+$0355						&	18:58:03.73	&	$+$03:55:08.04	&	3.45	&	55.58	&	31.5	&	29.6	&	1.6	&	1.1	&	146	&	6.55	&	...	\\
\textbf{J1208.4$-$6239}$^\dagger$	&	12:08:26.89	&	$-$62:39:26.13	&	1.56	&	64.44	&	39.2	&	52.0	&	4.9	&	1.8	&	874	&	6.43	&	...	\\
\textbf{J1350.4$-$6224}$^\dagger$	&	13:50:34.69	&	$-$62:23:43.53	&	1.71	&	58.24	&	41.3	&	90.8	&	2.4	&	0.7	&	357	&	6.41	&	...	\\
J1037.9$-$5843*						&	10:38:01.49	&	$-$58:44:20.62	&	4.29	&	38.88	&	24.9	&	163.9	&	0.4	&	0.0	&	391	&	6.32	&	...	\\
J2112.5$-$3044$^\dagger$			&	21:12:32.39	&	$-$30:43:58.53	&	1.39	&	51.84	&	69.0	&	151.0	&	2.8	&	1.1	&	1805&	6.25	&	...	\\
J1636.2$-$4709c						&	16:36:22.32	&	$-$47:09:53.05	&	4.41	&	57.44	&	13.7	&	4.4		&--		&	2.3	&	541	&	6.17	&	spp	\\
J1358.5$-$6025						&	13:58:24.20	&	$-$60:25:30.56	&	2.44	&	53.16	&	32.8	&	21.1	&	5.7	&	2.2	&	639	&	6.15	&	...	\\
J1048.2$-$5928						&	10:48:40.66	&	$-$59:26:03.43	&	3.98	&	65.78	&	101.1	&	60.4	&	1.5	&	1.4	&	381	&	6.11	&	...	\\
J2034.6$+$4302						&	20:34:58.42	&	$+$43:05:08.99	&	6.30	&	41.40	&	50.7	&	112.7	&	0.4	&	0.3	&	324	&	6.11	&	...	\\
J1754.0$-$2930$^\dagger$			&	17:54:14.33	&	$-$29:32:08.04	&	3.72	&	59.67	&	49.8	&	38.4	&	2.2	&	2.0	&	498	&	6.06	&	...	\\
J1214.0$-$6236$^\dagger$			&	12:14:10.04	&	$-$62:36:16.69	&	1.98	&	58.02	&	20.3	&	15.7	&	13.1&	2.2	&	789	&	6.05	&	spp	\\
J1652.8$-$4351						&	16:52:32.63	&	$-$43:56:50.10	&	6.40	&	64.55	&	31.0	&	62.0	&	1.3	&	0.9	&	184	&	6.00	&	...	\\
J1317.6$-$6315						&	13:17:35.62	&	$-$63:17:18.00	&	2.96	&	50.53	&	25.0	&	37.0	&	2.7	&	1.7	&	347	&	5.99	&	...	\\
J2039.4$+$4111						&	20:39:45.84	&	$+$41:09:34.39	&	3.61	&	45.39	&	48.2	&	98.1	&	0.3	&	0.3	&	249	&	5.91	&	...	\\
J1852.8$+$0158*						&	18:52:27.92	&	$+$02:01:37.54	&	4.17	&	54.52	&	12.1	&	0.2		&	--	&	2.8	&	838	&	5.89	&	...	\\
\textbf{J0631.6$+$0644}				&	06:31:49.76	&	$+$06:44:46.66	&	1.93	&	43.04	&	26.6	&	37.2	&	4.6	&	1.6	&	676	&	5.84	&	spp	\\
J1840.1$-$0412*						&	18:40:06.15	&	$-$04:11:35.22	&	2.95	&	30.14	&	15.9	&	0.0 	&	--	&	2.5	&	416	&	5.83	&	spp	\\
J1928.9$+$1739						&	19:29:02.93	&	$+$17:34:58.90	&	9.16	&	47.86	&	26.9	&	12.0	&	3.6	&	2.1	&	235	&	5.79	&	...	\\
J0225.8$+$6159						&	02:26:20.37	&	$+$62:00:10.48	&	3.49	&	46.69	&	28.8	&	29.7	&	2.2	&	1.7	&	473	&	5.77	&	...	\\
\textbf{J0002.6$+$6218}$^\dagger$	&	00:02:48.88	&	$+$62:16:54.71	&	2.25	&	48.02	&	58.0	&	80.3	&	1.8	&	1.5	&	716	&	5.76	&	...	\\
J1740.5$-$2642						&	17:40:41.52	&	$-$26:39:52.98	&	4.29	&	33.42	&	23.2	&	34.1	&	2.5	&	1.8	&	222	&	5.74	&	...	\\
J1834.5$-$0841*						&	18:34:31.66	&	$-$08:40:15.75	&	4.02	&	57.10	&	0.5		&	0.1		&	--	&	2.2	&	287	&	5.72	&	snr	\\
J2042.4$+$4209						&	20:42:39.77	&	$+$42:09:19.64	&	11.48	&	49.90	&	27.1	&	27.4	&	0.5	&	1.0	&	185	&	5.68	&	...	\\
J1814.0$-$1757c						&	18:13:24.52	&	$-$17:53:55.97	&	5.83	&	56.91	&	8.8		&	7.4		&	--	&	2.3	&	662	&	5.59	&	...	\\
J2041.1$+$4736$^\dagger$			&	20:41:08.34	&	$+$47:35:50.81	&	2.01	&	56.28	&	38.0	&	15.9	&	10.3&	2.3	&	967	&	5.53	&	...	\\
J1047.3$-$6005						&	10:47:21.66	&	$-$60:05:11.01	&	6.22	&	49.04	&	22.3	&	16.4	&	3.0	&	1.5	&	115	&	5.52	&	...	\\
J2039.6$-$5618						&	20:39:36.25	&	$-$56:17:12.94	&	1.82	&	34.60	&	30.4	&	60.3	&	3.9	&	1.6	&	1266&	5.47	&	...	\\
J1900.8$+$0337						&	19:00:37.96	&	$+$03:39:10.57	&	3.94	&	45.87	&	44.9	&	4.7		&	--	&	2.3	&	186	&	5.42	&	...	\\
J0855.4$-$4818						&	08:55:18.44	&	$-$48:14:13.02	&	10.69	&	33.84	&	53.0	&	66.4	&	0.5	&	0.9	&	288	&	5.39	&	...	\\
J1747.7$-$2904						&	17:47:51.94	&	$-$29:01:49.54	&	2.95	&	65.34	&	10.3	&	124.7	&	7.1	&	2.2	&	666	&	5.37	&	...	\\
J0541.1$+$3553						&	05:40:47.47	&	$+$35:54:40.72	&	8.53	&	35.17	&	37.3	&	33.0	&	1.8	&	1.9	&	329	&	5.34	&	...	\\
J1549.1$-$5347c*					&	15:48:38.12	&	$-$53:44:00.33	&	5.02	&	51.64	&	10.9	&	0.1		&	--	&	2.9	&	1172&	5.27	&	spp	\\
J1039.1$-$5809						&	10:38:25.85	&	$-$58:08:23.45	&	13.63	&	37.46	&	24.7	&	23.4	&	1.7	&	1.3	&	107	&	5.23	&	...	\\
J1831.7$-$0230						&	18:31:33.96	&	$-$02:31:25.54	&	5.83	&	31.11	&	17.8	&	2.1		&	--	&	2.7	&	421	&	5.23	&	...	\\
J1702.8$-$5656$^\dagger$			&	17:02:45.00	&	$-$56:54:39.46	&	1.88	&	58.78	&	46.9	&	53.1	&	3.4	&	2.1	&	1917&	5.19	&	...	\\
J1736.0$-$2701*						&	17:36:07.44	&	$-$27:03:29.55	&	6.88	&	38.45	&	25.2	&	23.7	&	0.3	&	0.0	&	80	&	5.18	&	...	\\
J2023.5$+$4126*						&	20:23:24.65	&	$+$41:27:31.08	&	4.35	&	48.95	&	78.1	&	36.7	&	0.4	&	0.0	&	93	&	5.12	&	...	\\
J1758.8$-$2346						&	17:59:09.58	&	$-$23:47:19.28	&	3.69	&	41.80	&	11.8	&	5.4		&	--	&	1.9	&	218	&	5.01	&	...	\\
J2004.4$+$3338*						&	20:04:22.03	&	$+$33:39:29.46	&	1.47	&	50.29	&	13.5	&	0.0		&	--	&	2.4	&	708	&	5.01	&	...	\\
J0212.1$+$5320						&	02:12:12.29	&	$+$53:20:49.61	&	1.58	&	51.47	&	45.9	&	82.0	&	3.3	&	1.5	&	1442&	5.01	&	...	\\
J1901.1$+$0728						&	19:01:09.32	&	$+$07:30:01.23	&	3.29	&	55.34	&	25.8	&	10.5	&	6.6	&	2.0	&	134	&	4.88	&	...	\\
J1503.5$-$5801						&	15:03:39.92	&	$-$58:00:43.22	&	3.88	&	67.48	&	26.3	&	18.7	&	3.7	&	2.0	&	359	&	4.85	&	...	\\
J1850.5$-$0024						&	18:50:31.56	&	$-$00:24:33.69	&	4.83	&	64.27	&	14.6	&	2.8		&	--	&	2.3	&	216	&	4.76	&	...	\\
J0933.9$-$6232$^\dagger$			&	09:34:00.41	&	$-$62:32:57.43	&	1.77	&	59.20	&	88.0	&	125.9	&	2.0	&	0.8	&	907	&	4.73	&	...	\\
J1620.0$-$5101						&	16:19:48.66	&	$-$51:00:57.34	&	4.03	&	50.48	&	9.7		&	1.0		&	--	&	2.1	&	121	&	4.72	&	...	\\
J1726.6$-$3530c						&	17:26:32.27	&	$-$35:33:37.61	&	5.18	&	60.31	&	11.9	&	1.8		&	--	&	2.6	&	335	&	4.67	&	...	\\
J1919.9$+$1407						&	19:20:11.19	&	$+$14:11:54.53	&	7.95	&	67.73	&	17.6	&	0.3		&	--	&	2.7	&	642	&	4.66	&	...	\\
J1119.9$-$2204$^\dagger$			&	11:19:59.45	&	$-$22:04:25.17	&	1.80	&	62.62	&	103.2	&	156.9	&	1.7	&	1.3	&	1949&	4.63	&	...	\\
J0907.0$-$4802*						&	09:07:18.05	&	$-$47:58:38.32	&	10.11	&	40.75	&	29.3	&	28.0	&	0.4	&	0.2	&	123	&	4.58	&	...	\\
J1718.0$-$3726						&	17:18:02.10	&	$-$37:26:50.06	&	1.02	&	41.58	&	1.5		&	2.0		&	--	&	2.1	&	593	&	4.55	&	snr	\\
J1859.6$+$0102						&	18:59:39.72	&	$+$01:00:15.56	&	5.43	&	68.61	&	18.9	&	13.1	&	3.5	&	1.8	&	150	&	4.40	&	...	\\
J2035.0$+$3634						&	20:35:02.11	&	$+$36:32:12.74	&	1.88	&	52.58	&	39.2	&	57.5	&	2.8	&	0.8	&	401	&	4.39	&	...	\\
J1345.1$-$6224						&	13:44:43.61	&	$-$62:28:30.64	&	5.12	&	58.30	&	12.8	&	1.3		&	--	&	2.7	&	568	&	4.39	&	spp	\\
J0744.1$-$2523						&	07:44:06.64	&	$-$25:25:17.47	&	1.97	&	61.34	&	40.9	&	55.3	&	3.2	&	1.8	&	666	&	4.27	&	...	\\
J0426.7$+$5437						&	04:26:33.79	&	$+$54:35:00.35	&	3.01	&	51.83	&	63.9	&	59.0	&	1.7	&	2.1	&	1235&	4.27	&	...	\\
J1539.2$-$3324$^\dagger$			&	15:39:20.23	&	$-$33:24:56.62	&	1.64	&	57.87	&	102.9	&	129.3	&	2.3	&	0.4	&	694	&	4.22	&	...	\\
J1641.1$-$4619c*					&	16:41:00.45	&	$-$46:19:46.25	&	1.87	&	39.43	&	0.7		&	0.2		&	--	&	2.3	&	292	&	4.15	&	spp	\\
\textbf{J1528.3$-$5836}				&	15:28:23.37	&	$-$58:38:05.98	&	1.87	&	68.72	&	44.9	&	41.4	&	4.0	&	1.6	&	452	&	4.14	&	...	\\
J1857.9$+$0355						&	18:58:03.73	&	$+$03:55:08.04	&	3.45	&	41.47	&	11.2	&	32.2	&	2.2	&	1.4	&	131	&	4.13	&	...	\\
J1855.4$+$0454						&	18:55:12.72	&	$+$04:55:38.38	&	4.46	&	38.60	&	6.6		&	4.4		&	--	&	2.4	&	193	&	4.12	&	...	\\
J1650.0$-$4438c*					&	16:49:48.42	&	$-$44:38:58.44	&	6.63	&	58.81	&	1.0		&	0.1		&	--	&	3.1	&	843	&	4.02	&	...	\\
J0901.6$-$4700						&	09:01:40.90	&	$-$46:52:10.77	&	7.02	&	55.10	&	30.0	&	52.7	&	1.0	&	1.2	&	221	&	4.02	&	...	\\
J1329.8$-$6109						&	13:29:57.92	&	$-$61:08:00.95	&	2.45	&	55.66	&	22.5	&	21.1	&	4.9	&	1.6	&	246	&	3.91	&	...	\\
J1639.4$-$5146						&	16:39:25.17	&	$-$51:46:04.03	&	1.39	&	58.03	&	4.2		&	2.8		&	--	&	2.3	&	945	&	3.85	&	...	\\
J1833.9$-$0711*						&	18:34:10.57	&	$-$07:11:34.47	&	3.12	&	82.07	&	1.6		&	0.4		&	--	&	2.3	&	482	&	3.85	&	spp	\\
J1814.1$-$1734c						&	18:14:07.87	&	$-$17:36:39.99	&	2.96	&	50.07	&	7.1		&	5.3		&	--	&	1.4	&	83	&	3.73	&	...	\\
J1139.0$-$6244						&	11:39:07.61	&	$-$62:46:04.02	&	2.31	&	29.45	&	7.5		&	16.5	&	8.6	&	1.9	&	278	&	3.71	&	...	\\
J1626.2$-$2428c						&	16:26:25.40	&	$-$24:31:36.54	&	4.74	&	46.87	&	15.9	&	7.8		&	--	&	2.1	&	392	&	3.66	&	...	\\
J1212.2$-$6251					&	12:12:18.06	&	$-$62:53:31.51	&	2.84	&	53.70	&	1.4		&	12.9	&	45.8&	2.4	&	426	&	3.45	&	spp	\\

\tablecomments{List of the 118 3FGL sources with log $R_{S}$ > 0 searched for gamma-ray pulsars using Einstein@Home, ranked by their probability to be pulsars according to the GMM analysis presented in Section~\ref{s:classification}. Sources marked with a $\dagger$ symbol were searched in a previous Einstein@Home \& \textit{Atlas} survey for gamma-ray pulsars. Sources for which suspiciously low or high cutoff energies were measured are marked with asterisks. We highlight in bold face the 3FGL sources in which pulsars were discovered in this survey. The discovery and analysis of PSRs J1906+0722 and J1208$-$6238 are presented in \citet{Clark2015ApJJ1906} and \citet{ClarkJ1208}, while PSRs J1035$-$6720 and PSR J1744$-$7619 discovered in 3FGL~J1035.7$-$6720 and 3FGL~J1744.1$-$7619 will be presented in a future publication. Columns 2 to 4 list the searched position and radius. Columns 5 and 6 give the variability index, VI, and curvature TS, $\textrm{TS}_\textrm{curve}$, from a preliminary version of the 3FGL catalog. Columns 7 to 10 give the TS of the spectral cutoff ($\textrm{TS}_\textrm{cut}$), the cutoff energy ($E_\textrm{cut}$), the photon index ($\Gamma$) and the source TS value from our binned maximum likelihood analysis with \texttt{Pointlike}; cutoff energies are listed for sources with $\textrm{TS}_\textrm{cut} > 9$. Column 11 lists the pulsar likelihood value from our GMM analysis. Column 12 lists association flags from the 3FGL catalog: ``pwn'' and ``snr'' labels indicate possible associations with pulsar wind nebulae (PWN) and supernova remnants (SNR) respectively, sources with class ``spp'' are special cases with potential PWN or SNR associations. Sources below the horizontal line were searched with relocalized positions, as mentioned in Sec. \ref{s:reloc}. }

\enddata
	
\end{deluxetable*}